\documentclass[letterpaper,journal]{IEEEtran}
\usepackage{amsmath,amsfonts}
\usepackage{amsthm}
\usepackage{algorithmic}
\usepackage{algorithm}
\usepackage{array}
\usepackage[caption=false,font=normalsize,labelfont=sf,textfont=sf]{subfig}
\usepackage{textcomp}
\usepackage{stfloats}
\usepackage{url}
\usepackage{verbatim}
\usepackage{graphicx}
\usepackage{booktabs}
\usepackage{cite}
\usepackage{orcidlink}
\usepackage{multirow}
\usepackage{hyperref}
\hypersetup{colorlinks=true, citecolor=blue, linkcolor=blue, urlcolor=black} 
\usepackage{xcolor}
\DeclareFontShape{OT1}{ptm}{m}{scit}{<->ssub * ptm/m/sc}{}
\DeclareFontShape{OT1}{ptm}{m}{sc}{<->ssub * ptm/m/n}{}
\DeclareFontShape{OT1}{ptm}{b}{sc}{<->ssub * ptm/b/n}{}
\DeclareFontShape{OT1}{ptm}{b}{scit}{<->ssub * ptm/b/sc}{}
\DeclareFontShape{OT1}{ptm}{bx}{n}{<->ssub * ptm/b/n}{}
\DeclareFontShape{OT1}{ptm}{bx}{it}{<->ssub * ptm/b/it}{}
\renewcommand{\eqref}[1]{\textup{\textcolor{blue}{\hyperref[#1]{(\ref*{#1})}}}}
\makeatletter
\let\orig@cite\@cite
\renewcommand{\@cite}[2]{\textcolor{blue}{\orig@cite{#1}{#2}}}
\makeatother

\usepackage{colortbl}
\usepackage{xcolor}
\definecolor{ac_gray}{gray}{.2}
\definecolor{ac_lightgray}{gray}{.65}
\definecolor{forestgreen}{RGB}{47, 159, 87}%
\definecolor{forestred}{RGB}{255,70,70}%

\begin{document}

\title{Circular Phase Representation and Geometry-Aware Optimization for Ptychographic Image Reconstruction}

\author{Carson Yu Liu$^{\orcidlink{0000-0002-7325-5763}}$
, Jun Cheng$^{\orcidlink{0000-0003-1786-6188}}$,~\IEEEmembership{Senior Member,~IEEE,} Chien-Chun Chen, Steve F. Shu$^{\orcidlink{0000-0003-1581-4405}}$,~\IEEEmembership{Senior Member,~IEEE}
\thanks{This work has been submitted to the IEEE for possible publication. Copyright may be transferred without notice, after which this version may no longer be accessible.}
\thanks{
Carson Yu Liu is with the School of Electrical and Computer Engineering, University of Sydney, Camperdown, NSW 2006, Australia (e-mail:
\href{mailto:carson.liu@sydney.edu.au}{carson.liu@sydney.edu.au}).

Jun Cheng is with the Institute for Infocomm Research, Agency for
Science, Technology and Research (A$\ast$STAR),, Singapore 138632 (e-mail:
\href{mailto:cheng_jun@a-star.edu.sg}{cheng\_jun@a-star.edu.sg}).

Chien-Chun Chen is with the Department of Engineering and System Science, National Tsing Hua University, Hsinchu 300044, Taiwan (e-mail:
\href{mailto:chenchienchun0627@gmail.com}{chenchienchun0627@gmail.com}).

Steve F. Shu is with the School of Electrical and Computer Engineering, University of Sydney, Camperdown, NSW 2006, Australia (e-mail:
\href{mailto:steve.shu@sydney.edu.au}{steve.shu@sydney.edu.au}).

Code will be available at: \url{https://github.com/carson-liu/CPR}.
}
\thanks{\textit{(Corresponding author: Steve F. Shu.)}}
}
\markboth{Journal of \LaTeX\ Class Files,~Vol.~14, No.~8, August~2021}%
{Shell \MakeLowercase{\textit{et al.}}: A Sample Article Using IEEEtran.cls for IEEE Journals}


\maketitle

\begin{abstract}
Traditional iterative reconstruction methods are accurate but computationally expensive, limiting their use in high-throughput and real-time ptychography. Recent deep learning approaches improve speed, but often predict phase as a Euclidean scalar despite its $2\pi$ periodicity, which can introduce wrapping artifacts, discontinuities at $\pm\pi$, and a mismatch between the loss and the underlying signal geometry.
We present a deep learning framework for ptychographic reconstruction that models phase on the unit circle using cosine and sine components. Phase error is optimized with a differentiable geodesic loss, which avoids branch-cut discontinuities and provides bounded gradients. The network further incorporates saturation-aware dual-gain input scaling, parallel encoder branches, and three decoders for amplitude, cosine, and sine prediction, together with a composite loss that promotes circular consistency and structural fidelity.
Experiments on synthetic and experimental datasets show consistent improvements in both amplitude and phase reconstruction over existing deep learning methods. Frequency-domain analysis further shows better preservation of mid- and high-frequency phase content. The proposed method also provides substantial speedup over iterative solvers while maintaining physically consistent reconstructions.
\end{abstract}


\section{Introduction}

\IEEEPARstart{P}{tychography} is a coherent diffraction imaging technique that reconstructs the complex-valued transmission function of a specimen from far-field diffraction patterns recorded at overlapping scan positions~\cite{Rodenburg2004}. By recovering both amplitude and phase, it enables high-resolution imaging and has been widely used in synchrotron materials science, biological imaging, and semiconductor metrology~\cite{Pfeiffer2017,Thibault2009}.

Conventional ptychographic reconstruction is typically based on iterative phase retrieval algorithms, such as the extended Ptychographic Iterative Engine (ePIE)~\cite{Maiden2009} and difference-map methods~\cite{Thibault2008}. Although physically grounded, these methods often require careful initialization, incur substantial computational cost, and may converge slowly or stagnate when data quality is limited.

Recent deep learning approaches aim to replace iterative solvers with feedforward networks that map diffraction patterns directly to amplitude and phase images, greatly reducing inference time~\cite{Cherukara2018,Cherukara2020,Chang2023,Pan2023,Yue2025}. Representative directions include multi-decoder convolutional architectures~\cite{Cherukara2018}, physics-informed objectives~\cite{Chang2023}, transfer learning strategies~\cite{Pan2023}, and attention mechanisms tailored to diffraction geometry~\cite{Yue2025}. These studies show that data-driven models can provide efficient alternatives in practical imaging settings. More broadly, recent learning-based phase retrieval methods have also considered low-dose or data-scarce regimes using implicit generative priors~\cite{Manekar2024}.

Despite this progress, phase representation remains a basic challenge in learning-based reconstruction. Phase is an angular quantity defined modulo $2\pi$, and handling wrapped phase has long been a central issue in phase imaging and phase unwrapping problems~\cite{BioucasDias2007, Yang2021}. Most existing methods, however, predict phase directly as a scalar and optimize it with pixel-wise Euclidean losses. This treats phase as a linear variable and introduces an artificial discontinuity at the wrapping boundary.

The mismatch between circular phase geometry and linear regression can degrade optimization. Phase values that are close on the circle, especially near $\pm\pi$, may appear far apart under a Euclidean metric, leading to misleading gradients. Wrapping the phase difference corrects the distance at the value level, but does not remove the discontinuity in the representation itself.

In this work, we address phase reconstruction from the perspective of representation and optimization. We embed phase on the unit circle and define learning objectives directly on this manifold, aligning the loss with the geometry of the signal. This avoids explicit phase wrapping and yields well-behaved gradients over the full angular domain.

Specifically, this paper makes the following contributions:
\begin{itemize}
\item We introduce a circular phase representation in which the network predicts the cosine and sine components of phase, removing the artificial discontinuity of scalar phase regression. Phase is recovered with a numerically stable $\operatorname{atan2}$ operation.

\item We design a composite loss on the circular manifold that combines geodesic angular discrepancy, a unit-circle consistency constraint, spatial gradient regularization, and structural similarity terms.

\item We develop a saturation-aware dual-gain architecture with parallel encoder branches, a shared learnable skip connection, and dedicated decoders for amplitude and circular phase components, enabling robust reconstruction across a wide dynamic range.
\end{itemize}

Extensive ablation studies further verify the importance of the circular representation and its associated loss terms for reconstruction accuracy and stability.

\section{Related Work}

Ptychography has traditionally relied on iterative reconstruction, in which the complex object $O(\mathbf r)$ and often the probe $P(\mathbf r)$ are jointly estimated by enforcing Fourier-domain intensity consistency and real-space overlap constraints. Representative methods include the extended ptychographic iterative engine (ePIE), which performs sequential object--probe updates at each scan position~\cite{Maiden2009}, and the difference map, which formulates phase retrieval as alternating projections between constraint sets~\cite{Elser2003}. These iterative solvers remain standard baselines in optical and electron ptychography, but they usually require many iterations with repeated FFTs and full data passes, and their convergence can be sensitive to initialization, noise, and acquisition conditions~\cite{Wang2024,Dong2023}. These limitations motivate learning-based alternatives for scalable and real-time reconstruction.

Deep learning reformulates ptychographic phase retrieval as a feed-forward prediction problem, enabling direct estimation of amplitude and phase from diffraction measurements. Early studies showed that convolutional networks can achieve high-quality reconstruction with substantially lower inference cost~\cite{Guan2019}, including under experimental noise without iterative refinement~\cite{Wengrowicz2020}. More recent work increasingly incorporates physics and acquisition structure into the model design. Deep iterative phase retrieval inserts a neural module into each iteration to refine intermediate estimates~\cite{DeepIterative2022}, while Yamada \textit{et al.} combine a denoising network with a physics-driven solver to improve robustness under varying acquisition conditions~\cite{Yamada2024}. Architectural adaptations have also been explored: PPN introduces polar coordinate attention to capture radial and angular correlations~\cite{Yue2025}, PtychoFormer adopts a hierarchical transformer with hybrid refinement to reduce global phase ambiguity while preserving speed~\cite{Nakahata2024}, and system-level acceleration has been demonstrated in AI-enabled coherent diffraction imaging and real-time edge deployment~\cite{Cherukara2020,Babu2023}.

Despite these advances, most deep ptychographic models still predict phase as a scalar field and optimize pixel-wise Euclidean losses~\cite{Cherukara2018,Guan2019,Cherukara2020,Wengrowicz2020,Chang2023,Pan2023,Nakahata2024,Yue2025,Wang2024,Dong2023}. More broadly, prior work in learning with angular variables shows that sine--cosine parameterizations can remove boundary discontinuities by embedding the unit circle into $\mathbb{R}^2$~\cite{Beyer2015}, and manifold-aware regression further suggests that respecting the underlying spherical, and by extension circular, geometry can improve stability for variables defined on closed manifolds~\cite{Liao2019}. Motivated by these observations, our method models phase in circular coordinates and introduces geometry-aware supervision to better match the periodic structure of the target variable.

\section{Methodology}
\label{sec:method}

\subsection{Problem Formulation}
\label{sec:problem}

We consider a ptychographic experiment in which a coherent probe $P(\mathbf{r})$ illuminates a specimen described by a complex transmission function $O(\mathbf{r})$ at a set of overlapping scan positions $\{\mathbf{r}_j\}_{j=1}^{N}$.
At the $j$-th scan position, the shifted probe multiplicatively modulates the specimen, producing the exit wave $\psi_j(\mathbf{r}) = P(\mathbf{r}-\mathbf{r}_j)\, O(\mathbf{r})$, where $\mathbf{r}$ denotes the spatial coordinate in the object plane.
The corresponding far-field diffraction intensity recorded on the detector is
\begin{equation}
\label{eq:diffraction}
I_j(\mathbf{q})=\left|\mathcal{F}\{\psi_j(\mathbf{r})\}\right|^2
\end{equation}
where $\mathcal{F}(\cdot)$ denotes the Fourier transform and $\mathbf{q}$ denotes the reciprocal-space coordinate.
Equation~\eqref{eq:diffraction} captures the key ambiguity in ptychography: only the magnitude of the Fourier transform is observed, while the Fourier phase is not directly measured.

The object transmission function can be written in polar form as
\begin{equation}
O(\mathbf{r}) = A(\mathbf{r})\,\exp \!\bigl(i\,\phi(\mathbf{r})\bigr)
\end{equation}
where $A(\mathbf{r})$ denotes the amplitude and $\phi(\mathbf{r})$ denotes the phase.
The amplitude reflects attenuation (or gain) of the wavefield, and the phase encodes optical path-length variations induced by the specimen.

Given diffraction intensities $\{I_j\}_{j=1}^{N}$, the goal of deep learning--based ptychographic reconstruction is to learn a parameterized mapping $f_{\theta}: I_j \mapsto (\hat{A}_j,\hat{\phi}_j)$, which predicts the amplitude $\hat{A}_j$ and phase $\hat{\phi}_j$ associated with the $j$-th diffraction intensity measurement $I_j$.

\begin{figure*}[h!]
\centering
\includegraphics[width=0.95\linewidth]{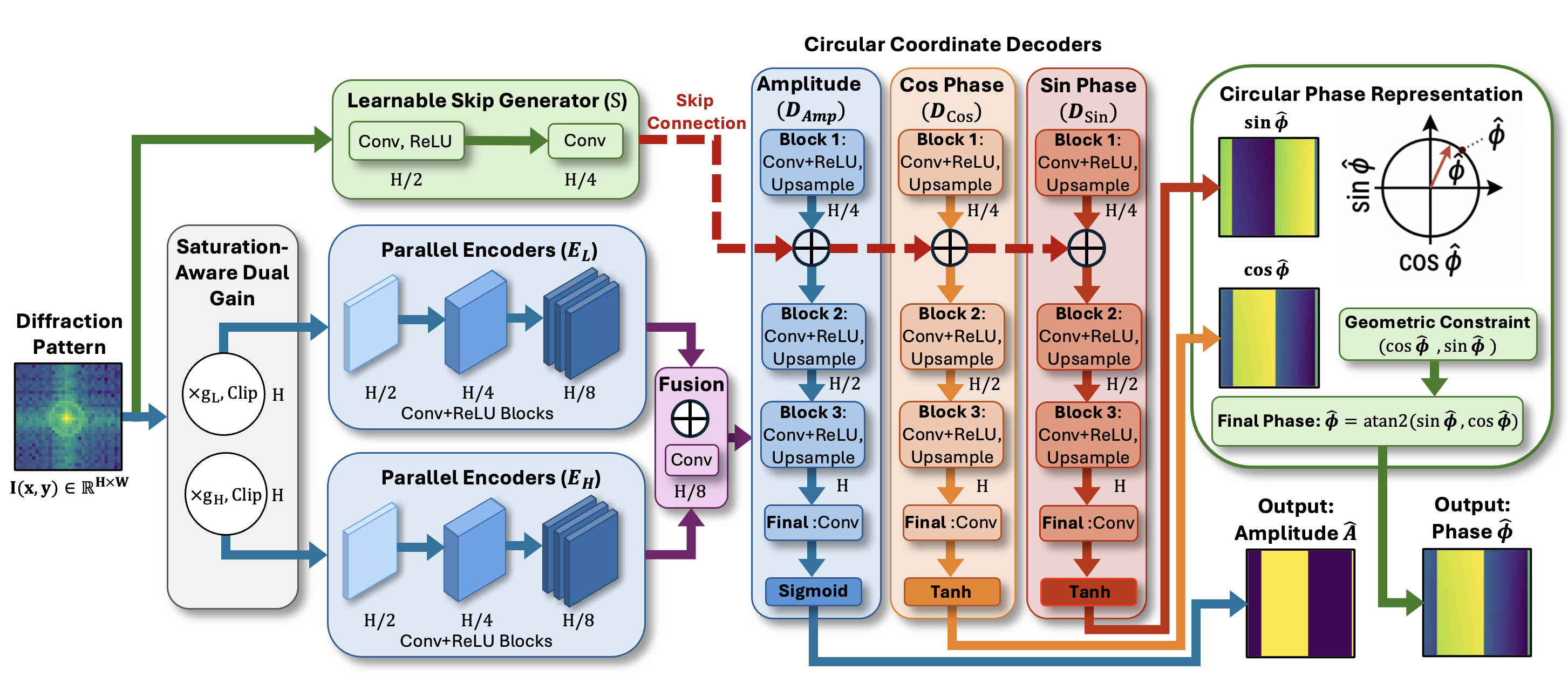}
\caption{Overview of the proposed network. The input diffraction intensity $I$ is transformed into low-gain and high-gain views $(I_l,I_h)$ via saturation-aware dual-gain scaling. Two encoder branches extract features from $(I_l,I_h)$, which are fused by channel projection. A shared mid-resolution skip feature is computed from the original input and injected into each decoder. Three decoders predict amplitude $\hat{A}$ and circular phase coordinates $(\hat{c},\hat{s})$. Unit-circle projection (Eq.~\eqref{eq:unit_norm}) produces normalized coordinates $(\tilde{c},\tilde{s})$, and phase is recovered by $\hat{\phi}=\mathrm{atan2}(\tilde{s},\tilde{c})$.}
\label{fig:architecture}
\end{figure*}

\subsection{Circular Phase Representation and Optimization}
\label{sec:phase_representation}

This subsection specifies the phase output representation and the associated training objectives.
Since phase is an angular variable, a scalar parameterization combined with Euclidean losses can introduce a branch-cut discontinuity and mismatched gradients during optimization.
We therefore use a circular representation in which phase is embedded on the unit circle, and losses are defined to be consistent with this geometry.

\subsubsection{Phase as a Circular Quantity}
\label{sec:phase_circular}

Phase is defined modulo $2\pi$, so values differing by integer multiples of $2\pi$ represent the same physical state.
The phase domain is therefore topologically equivalent to the unit circle $S^1$.
In learning-based reconstruction, this motivates output representations and losses that respect periodicity without introducing an artificial discontinuity.

\subsubsection{Topological Inconsistency of Scalar Phase Regression}
\label{sec:topological_inconsistency}

Many learning-based approaches predict phase as a scalar and optimize pixel-wise Euclidean losses.
This imposes a linear structure on a circular variable and introduces an artificial branch cut.
As a result, phase values that are adjacent on $S^1$ may appear far apart under Euclidean distance, which can distort gradient signals.

To illustrate, consider two phase values located on opposite sides of the branch cut, $\phi_1 = -\pi + \epsilon$ and $\phi_2 = \pi - \epsilon$, where $\epsilon > 0$ is small.
On $S^1$, their geodesic (shortest-arc) distance is only $\Delta \phi_{\mathrm{geo}} = 2\epsilon$, which vanishes as $\epsilon \to 0$.
In contrast, treating phase as a scalar gives a Euclidean discrepancy of $|\phi_1 - \phi_2| = 2\pi - 2\epsilon$, which approaches the maximal separation on the circle.
When an $L_2$ loss is applied to scalar phase values, this mismatch can produce large gradients even when predictions are close in angular distance.

A common remedy is to wrap the phase difference before computing the loss so that the value-level discrepancy reflects the shortest angular distance.
However, the wrapping operator is discontinuous and non-differentiable at the branch cut, leading to undefined or unstable gradients precisely where corrective updates are most needed.
Wrapping therefore corrects the distance computation at the value level, but does not provide a smooth optimization objective.

Table~\ref{tab:phase_comparison} summarizes the implications of these choices in terms of distance and gradient behavior.

\begin{table}[!ht]
\centering
\caption{Comparison of phase regression strategies.}
\label{tab:phase_comparison}
\resizebox{\linewidth}{!}{%
\begin{tabular}{@{}lcccc@{}}
\toprule
\textbf{Method} 
& \textbf{Correct Dist.} 
& \textbf{Smooth $\nabla$} 
& \textbf{Bounded $\nabla$} 
& \textbf{Topology-Aware} \\
\midrule
Scalar \\ + L2 
& $\times$ 
& \checkmark 
& $\times$ 
& $\times$ \\
Wrapped \\ + L2 
& \checkmark 
& $\times$ 
& $\times$ 
& $\times$ \\
Circular coords. \\ + $L_{\mathrm{circ}}$ (Ours)
& \checkmark 
& \checkmark 
& \checkmark 
& \checkmark \\
\bottomrule
\end{tabular}%
}
\vspace{0.5mm}
\footnotesize\emph{Note:} ``Correct Dist.'' uses the shortest-arc distance on $S^1$; $\nabla$ is taken with respect to the phase prediction.
\end{table}

\subsubsection{Circular Embedding and Output Parameterization}
\label{sec:output_parameterization}

To avoid introducing a branch cut in the output space, we represent phase using the embedding $\phi \mapsto (\cos\phi, \sin\phi)$, which maps $S^1$ into $\mathbb{R}^2$.
Accordingly, the network predicts two channels, denoted by $(\hat{c}, \hat{s})$, intended to approximate $(\cos\phi, \sin\phi)$.

We then apply an explicit unit-circle projection to obtain normalized circular coordinates:
\begin{equation}
\label{eq:unit_norm}
(\tilde{c}_i,\tilde{s}_i)
=
\frac{(\hat{c}_i,\hat{s}_i)}{\sqrt{\hat{c}_i^2+\hat{s}_i^2+\varepsilon}}
\end{equation}
where $i$ indexes spatial locations and $\varepsilon$ is a small constant for numerical stability.
The phase can then be recovered as $\hat{\phi} = \mathrm{atan2}(\tilde{s}, \tilde{c})$.

\subsubsection{Differentiable Circular Loss}
\label{sec:circular_loss}

Given the circular representation, we measure angular discrepancy using the periodic loss
\begin{equation}
L_{\text{circular}}(\phi_{\text{true}}, \phi_{\text{pred}})
= 1 - \cos(\phi_{\text{pred}} - \phi_{\text{true}})
\end{equation}
This loss is non-negative, periodic, and bounded, and it yields smooth gradients over all angles.
Because it depends only on the relative angle, it is naturally consistent with the geometry of $S^1$ and avoids the discontinuity issues that arise in scalar phase regression.

\subsection{Network Architecture}
\label{sec:arch_overview}

Building on the circular output parameterization in Section~\ref{sec:output_parameterization}, we design a network with dedicated decoders for amplitude and circular phase components.
Fig.~\ref{fig:architecture} provides an overview.

\subsubsection{Saturation-Aware Dual-Gain Scaling}
\label{sec:sadgs}

Diffraction patterns in ptychography often contain strong low-frequency peaks that may exceed the detector saturation threshold, while weak high-angle fringes carrying high-frequency structural information lie close to the noise floor.
To address this dynamic-range imbalance, we introduce a Saturation-Aware Dual-Gain Scaling (SADGS) module as the first processing stage.

Given raw diffraction intensity $I$ and two gain exponents $g_l$ and $g_h$ with $g_l < g_h$, the module produces two scaled branches:
\begin{equation}
\label{eq:dual_gain}
I_l = \min(k_l I,\; I_{\mathrm{sat}}), \qquad
I_h = \min(k_h I,\; I_{\mathrm{sat}})
\end{equation}
where $I_{\mathrm{sat}}$ is the detector saturation threshold, set to 4095 for a 12-bit detector.
The gain factor is defined as $k = \frac{I_{\mathrm{sat}}\alpha}{2^{g_0} I_{\max}} \cdot 2^{g_k}$ for $g_k \in \{g_l, g_h\}$, where $\alpha = 0.85$ is a headroom factor, $g_0$ is a base gain, and $I_{\max}$ is estimated from the training set as the 99.5th percentile of the intensity distribution and then fixed.
In practice, $g_l = 0.001$ suppresses clipping in strong peaks, whereas $g_h = 4$ enhances weak fringes.

\subsubsection{Dual-Branch Encoder and Learned Fusion}
\label{sec:encoder}

Each scaled input, $I_l$ and $I_h$, is processed by an independent encoder branch.
Formally, each encoder $E_k$, with $k \in \{l,h\}$, maps $\mathbb{R}^{1\times H\times W}$ to $\mathbb{R}^{4n_c\times \frac{H}{8}\times \frac{W}{8}}$.
Each encoder contains three convolutional stages with channel widths $[n_c, 2n_c, 4n_c]$, where $n_c=32$, using $5{\times}5$ kernels, stride-2 downsampling, and ReLU activations.

The encoder outputs $\mathbf{z}_l = E_l(I_l)$ and $\mathbf{z}_h = E_h(I_h)$ are concatenated along channels and projected back to the shared bottleneck width through a bias-free $1{\times}1$ convolution, yielding $\mathbf{z} = W_{1\times 1}[\mathbf{z}_l;\mathbf{z}_h]$, where $W_{1\times 1} \in \mathbb{R}^{4n_c\times 8n_c\times 1\times 1}$.
This linear projection enables data-driven fusion of the two dynamic-range views without adding spatial receptive field or bias terms.
As shown in the ablation study (Section~\ref{sec:ablation}), replacing this simple fusion with deeper convolutional fusion does not improve performance.

A separate learnable skip connection is generated from the raw input using two stride-2 $3{\times}3$ convolutions, with an intermediate ReLU after the first layer, producing $\mathbf{s} \in \mathbb{R}^{2n_c\times \frac{H}{4}\times \frac{W}{4}}$.
This shared skip path is fed to all three decoders and provides mid-resolution structural context during upsampling.

\subsubsection{Circular Coordinate Decoders}
\label{sec:decoders}

Three independent decoder branches, $\mathcal{D}_{Amp}$, $\mathcal{D}_{Cos}$, and $\mathcal{D}_{Sin}$, reconstruct amplitude $\hat{A}$, cosine component $\hat{c}$, and sine component $\hat{s}$, respectively.
All three decoders share the same architecture and differ only in their output activation and target quantity.

Each decoder contains three upsampling stages with convolutional refinement.
Starting from the bottleneck feature map of size $(4n_c,\frac{H}{8}\times\frac{W}{8})$, Block 1 applies two $3{\times}3$ convolutions with ReLU and upsamples to resolution $\frac{H}{4}\times\frac{W}{4}$.
In Block 2, the upsampled features are concatenated with the shared skip connection $\mathbf{s}$, giving $4n_c+2n_c$ input channels; two $3{\times}3$ convolutions reduce this to $2n_c$, followed by $2{\times}$ bilinear upsampling.
Block 3 applies two further $3{\times}3$ convolutions at width $2n_c$ and upsamples to the full resolution $H\times W$.

A final $3{\times}3$ convolution maps decoder features to one output channel.
The amplitude branch uses a sigmoid activation, while the cosine and sine branches use $\tanh$, which bounds the preprojection phase coordinates before the unit-circle normalization in Eq.~\eqref{eq:unit_norm}.
Using separate decoders rather than a shared multi-head output allows amplitude and phase components to follow distinct feature transformation paths.

\subsection{Amplitude--Phase Reconstruction Loss with Circular Consistency}
\label{sec:loss}

Accurate amplitude and phase reconstruction requires balancing pixel-wise fidelity, structural similarity, high-frequency preservation, and geometric consistency. We therefore use a composite loss in which each term addresses a distinct failure mode observed in preliminary experiments.

\subsubsection{Base Term: Pixel-wise Fidelity}

The base loss provides the core reconstruction objective:
\begin{equation}
\label{eq:base_loss}
\mathcal{L}_{\mathrm{base}} = \ell(A,\hat{A}) + \ell(c,\hat{c}) + \ell(s,\hat{s})
\end{equation}
where $\ell(X,\hat{X})$ is the mean squared error, i.e., the average of $(X_i-\hat{X}_i)^2$ over all pixels.

We use MSE because it provides smooth, stable gradients and emphasizes larger reconstruction errors. In diffraction-based reconstruction, such errors often correspond to meaningful physical mismatches rather than pure noise, making MSE more effective than MAE during early-stage optimization.

\subsubsection{Amplitude Term: Balancing Accuracy and Structure}

To improve amplitude reconstruction beyond pixel-wise fidelity, we add gradient and structural similarity terms:
\begin{equation}
\label{eq:amp_loss}
\mathcal{L}_{\mathrm{amp}}
=
\underbrace{\lambda_g \mathcal{L}_{\nabla}(A,\hat{A})}_{\text{gradient preservation}}
+
\underbrace{\lambda_s \mathcal{L}_{s}(A,\hat{A})}_{\text{structural similarity}}
\end{equation}

The gradient term $\mathcal{L}_{\nabla}$ computes, over all $N$ pixels, the average of the absolute differences between the horizontal gradients $(\nabla_x X,\nabla_x \hat{X})$ and the vertical gradients $(\nabla_y X,\nabla_y \hat{X})$.
This helps preserve sharp edges and fine structures that pixel-wise losses alone tend to smooth out. Empirically, $\lambda_g=0.12$ provides a good balance between edge preservation and global amplitude accuracy.

The structural term $\mathcal{L}_s$ is defined as $1-\mathrm{SSIM}(X,\hat{X})$, computed with a Gaussian window ($\sigma=1.5$, window size $11$) and standard constants $C_1=(0.01)^2$ and $C_2=(0.03)^2$. This term improves local luminance, contrast, and structural consistency, and with $\lambda_s=0.1$ acts as a useful regularizer without dominating optimization.

\subsubsection{Phase Term: Circular Representation with Geometric Consistency}

Phase reconstruction is more challenging because of periodicity and wrapping. Our framework predicts $(\cos\phi,\sin\phi)$ so that phase lies on the unit circle $S^1$, and the phase loss combines gradient preservation, structural similarity, and circular geodesic consistency:
\begin{equation}
\label{eq:phase_loss}
\begin{split}
\mathcal{L}_{\mathrm{phase}} =\;
& \underbrace{\lambda_g\bigl(\mathcal{L}_{\nabla}(c,\hat{c})+\mathcal{L}_{\nabla}(s,\hat{s})\bigr)}_{\text{gradient preservation}}
+ \underbrace{\lambda_s\bigl(\mathcal{L}_{s}(c,\hat{c})+\mathcal{L}_{s}(s,\hat{s})\bigr)}_{\text{structural similarity}} \\
& + \underbrace{\lambda_{\mathrm{circ}}\,\mathcal{L}_{\mathrm{circular}}}_{\text{geodesic distance}}
\end{split}
\end{equation}
where $c=\cos\phi$ and $s=\sin\phi$ are the ground-truth circular coordinates.

As in the amplitude branch, the gradient and SSIM terms preserve local structure in both circular channels. However, Euclidean similarity in $(c,s)$ space does not fully capture angular discrepancy on $S^1$, so we further introduce a circular geodesic loss:
\begin{equation}
\label{eq:circular_loss}
\mathcal{L}_{\mathrm{circular}}
=
\frac{1}{N}\sum_{i=1}^{N}\left[1-(c_i\hat{c}_i+s_i\hat{s}_i)\right]
=
\frac{1}{N}\sum_{i=1}^{N}\left[1-\cos(\Delta\phi_i)\right]
\end{equation}
where $\Delta\phi_i=\phi_i-\hat{\phi}_i$ is the angular error at pixel $i$.

This loss is periodic, smooth, and aligned with the shortest-arc discrepancy on $S^1$. For small angular errors, it behaves like a quadratic penalty since $1-\cos(\Delta\phi)\approx (\Delta\phi)^2/2$, which supports stable refinement near the optimum. It is also globally bounded, reaching $2$ at $\Delta\phi=\pi$, so large errors do not produce unbounded gradients. Its gradient is proportional to $\sin(\Delta\phi_i)$ and remains smooth across phase-wrapping boundaries, avoiding the discontinuities of direct scalar phase regression.

\subsubsection{Manifold Consistency Constraint}

To regularize the circular representation before projection, we impose a consistency loss that penalizes deviations from the unit-circle manifold:
\begin{equation}
\label{eq:cons_loss}
\mathcal{L}_{\mathrm{cons}} =
\frac{1}{N}\sum_{i=1}^{N}\bigl(\hat{c}_i^2+\hat{s}_i^2-1\bigr)^2
\end{equation}

Although Eq.~\eqref{eq:unit_norm} enforces the unit-circle constraint at the output, this term encourages pre-normalized predictions to remain close to the manifold, improving optimization stability and regularizing the learned circular representation.

\subsubsection{Overall Training Objective}

The final training objective combines the base, amplitude, phase, and consistency terms:
\begin{equation}
\label{eq:total_loss}
\mathcal{L}_{\mathrm{total}}
=
w_b\,\mathcal{L}_{\mathrm{base}}
+
w_a\,\mathcal{L}_{\mathrm{amp}}
+
w_p\,\mathcal{L}_{\mathrm{phase}}
+
w_c\,\mathcal{L}_{\mathrm{cons}}
\end{equation}

All models are implemented in PyTorch Lightning~\cite{Falcon2019PL} for standardized training and reproducible experiment management. Loss weights are fixed across experiments and selected on a held-out validation set by first matching the typical magnitudes of individual terms, then refining them through a small local search. Unless otherwise stated, we use $w_b=1.0$, $w_a=1.0$, $w_p=1.3$, $w_c=0.1$, $\lambda_{\mathrm{circ}}=0.6$, $\lambda_g=0.12$, and $\lambda_s=0.1$.

Accurate amplitude and phase reconstruction requires balancing pixel-wise fidelity, structural similarity, high-frequency preservation, and geometric consistency. We therefore use a composite loss in which each term addresses a distinct failure mode observed in preliminary experiments.

\section{Experiments And Results}

\begin{figure*}[htbp!]
\centering
\includegraphics[width=0.8\linewidth]{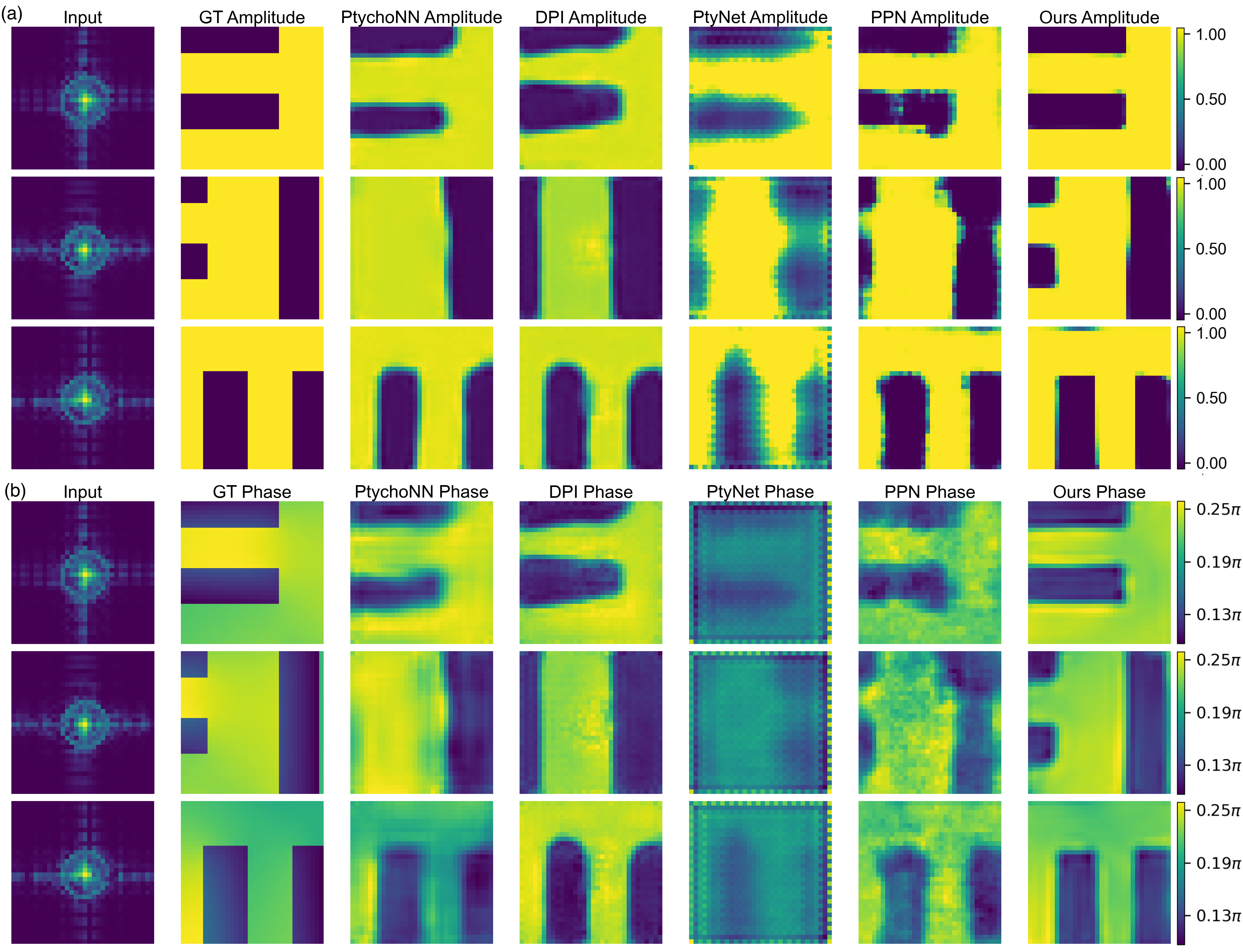}
\caption{Comparative evaluation of single-shot amplitude and phase reconstructions.
(a) Amplitude and (b) phase reconstruction results across five deep learning–based methods and the ground truth for three representative test samples.
Columns show, in order, the input diffraction pattern, ground truth, PtychoNN, DPI, PtyNet, PPN, and the proposed method.
All results are displayed using consistent color scales within each modality to facilitate fair visual comparison.}
\label{fig_single}
\end{figure*}

\subsection{Data Preparation and Partitioning}

We train and evaluate on two complementary datasets.
The first is a small synthetic dataset on a $61{\times}61$ scan grid (3721 samples), used for controlled experiments and rapid ablations.
The second is a much larger real experimental dataset from PtychoNN~\cite{Cherukara2020} on a $161{\times}161$ grid (25{,}921 samples), used to assess scalability, robustness to experimental effects, and generalization to a different diffraction input resolution ($64{\times}64$).

\textit{Synthetic LTEM dataset.}
The synthetic data are generated from a slightly modified 1951 USAF resolution chart from AD\_LTEM~\cite{Zhou2021}.
Using a $32{\times}32$ probe with $75\%$ overlap and random 3-pixel positional jitter, we simulate diffraction measurements on a $61{\times}61$ grid.
Poisson and Gaussian noise are added to model photon statistics and detector readout effects, following~\cite{Yue2025}.
For partitioning, we use a contiguous row-wise split: 49 scan rows for training, with $5\%$ of the training samples held out for validation, and the remaining 12 rows for testing.

\textit{Real experimental X-ray dataset.}
We also evaluate on the real X-ray ptychography dataset in~\cite{Cherukara2020}, acquired at the Advanced Photon Source (sector 26) from a tungsten test pattern with random etched features.
The scan forms a $161{\times}161$ grid with 30\,nm step size and about $50\%$ overlap, and diffraction patterns are recorded by a Medipix3 detector.
We follow the original split: the first 100 scan lines are used for training and validation, and the remaining 61 lines for testing.
The diffraction frames are resized to $64{\times}64$ before being fed into the network.

\subsection{Training Strategy}
\begin{algorithm}[t]
\caption{Algorithm 1: Training with Circular Phase Coordinates}
\label{alg:training}
\begin{algorithmic}[1]
\REQUIRE Training set $\{(I_j, A_j, \phi_j)\}_{j=1}^{M}$; loss weights $w_b, w_a, w_p, w_c, \lambda_c, \lambda_g, \lambda_s$
\ENSURE Trained model parameters $\theta^*$

\STATE \textbf{// Pre-compute circular coordinates}
\FOR{$j = 1$ \TO $M$}
    \STATE $c_j \leftarrow \cos(\phi_j)$; \quad $s_j \leftarrow \sin(\phi_j)$
\ENDFOR

\STATE Initialize model parameters $\theta$; best loss $\leftarrow +\infty$
\STATE Configure Adam optimizer with CyclicLR schedule

\FOR{epoch $= 1$ \TO $N_{\mathrm{epochs}}$}
    \FOR{each mini-batch $(I, A, c, s)$}
        \STATE \textbf{// Forward pass (proposed network)}
        \STATE $\hat{A}, \hat{c}, \hat{s} \leftarrow f_\theta(I)$
        \hfill{\scriptsize$\triangleright$ includes unit-circle norm.}

        \STATE \textbf{// Total physics-informed loss}
        \STATE $\mathcal{L}_{\mathrm{base}} \leftarrow $ Eq.~\eqref{eq:base_loss}; $\mathcal{L}_{\mathrm{amp}} \leftarrow $ Eq.~\eqref{eq:amp_loss}
        \STATE $\mathcal{L}_{\mathrm{phase}} \leftarrow $ Eq.~\eqref{eq:phase_loss}; $\mathcal{L}_{\mathrm{cons}} \leftarrow $ Eq.~\eqref{eq:cons_loss}
        \STATE $\mathcal{L}_{\mathrm{total}} \leftarrow w_b \mathcal{L}_{\mathrm{base}} + w_a \mathcal{L}_{\mathrm{amp}} + w_p \mathcal{L}_{\mathrm{phase}} + w_c \mathcal{L}_{\mathrm{cons}}$

        \STATE \textbf{// Update with gradient clipping}
        \STATE $\theta \leftarrow \text{Adam}(\nabla_\theta \mathcal{L}_{\mathrm{total}})$ with $\|\nabla\|_{\max}=1.0$
        \STATE Update CyclicLR schedule
    \ENDFOR

    \STATE \textbf{// Model selection on validation set}
    \STATE $\mathcal{L}_{\mathrm{val}} \leftarrow$ evaluate on validation set
    \IF{$\mathcal{L}_{\mathrm{val}} < $ best loss}
        \STATE best loss $\leftarrow \mathcal{L}_{\mathrm{val}}$; save $\theta$
    \ENDIF
\ENDFOR

\RETURN $\theta^* \leftarrow$ best saved parameters
\end{algorithmic}
\end{algorithm}

The Adam optimizer~\cite{kingma2015adam} is employed with an initial learning rate $\eta=10^{-3}$.
A triangular-2 cyclic learning rate schedule~\cite{Smith2017} oscillates between $\eta/10$ and $\eta$ with a half-cycle length of six epochs (computed as $6\times$ the number of iterations per epoch).
The decaying maximum in the triangular-2 mode provides an annealing effect that stabilizes late-stage training.
Gradient norms are clipped to a maximum of 1.0 to prevent training instabilities arising from the multi-component loss function. Algorithm \ref{alg:training} details end-to-
end optimization with circular phase coordinates and physical
constraints.

\subsection{Evaluation Metrics}

The proposed method is compared against four recent deep learning--based ptychographic reconstruction methods: PtychoNN~\cite{Cherukara2020}, DPI~\cite{Chang2023}, PtyNet~\cite{Pan2023}, and PPN~\cite{Yue2025}.
All methods are evaluated on the same set of 144 test samples from synthetic dataset using four standard image quality metrics: MSE, MAE, peak signal-to-noise ratio (PSNR), and SSIM.
Metrics are computed separately for amplitude and phase reconstructions.
Two evaluation modes are considered: (i) \emph{per-sample evaluation}, where metrics are computed for each reconstructed patch independently and summarized as population statistics; and (ii) \emph{stitched full-field evaluation}, where individually reconstructed patches are assembled into a large-field-of-view image using adaptive weighted stitching with a radially decaying quadratic weight kernel, and metrics are computed on the resulting composite.

\begin{table*}[htbp!]
    \caption{Reconstruction metrics for per-sample evaluation under two conditions: \textit{Noise-Free}
             and \textit{With Noise}. 
                 MSE ($\times 10^{-2}$), MAE ($\times 10^{-1}$), PSNR (dB), SSIM (\%).}
    \label{tab:combined_metrics}
    \centering
    \resizebox{1\textwidth}{!}{
    \begin{tabular}{@{\hspace{2mm}}lcccccccc@{\hspace{2mm}}}
        \toprule
        \multirow{2}{*}{\textcolor{ac_gray}{Methods}} 
        & \multicolumn{4}{c}{\textcolor{ac_gray}{Amplitude}} 
        & \multicolumn{4}{c}{\textcolor{ac_gray}{Phase}} \\
        \cmidrule(lr){2-5} \cmidrule(lr){6-9}
        & \textcolor{ac_gray}{MSE ($\times 10^{-2}$) $\downarrow$} 
        & \textcolor{ac_gray}{MAE ($\times 10^{-1}$) $\downarrow$} 
        & \textcolor{ac_gray}{PSNR (dB) $\uparrow$} 
        & \textcolor{ac_gray}{SSIM (\%) $\uparrow$}
        & \textcolor{ac_gray}{MSE ($\times 10^{-2}$) $\downarrow$} 
        & \textcolor{ac_gray}{MAE ($\times 10^{-1}$) $\downarrow$} 
        & \textcolor{ac_gray}{PSNR (dB) $\uparrow$} 
        & \textcolor{ac_gray}{SSIM (\%) $\uparrow$} \\
        \midrule
        \multicolumn{9}{l}{\textbf{Noise-Free}} \\
        \midrule
        PtychoNN 
        & $7.92 \pm 6.75$  & $1.04 \pm 0.69$ & $12.06 \pm 2.96$ & $65.07 \pm 16.13$
        & $1.44 \pm 1.58$  & $0.83 \pm 0.43$ & $16.24 \pm 3.48$ & $56.78 \pm 14.47$ \\
        DPI      
        & $7.11 \pm 5.94$  & $1.11 \pm 0.60$ & $12.44 \pm 2.78$ & $59.41 \pm 14.21$
        & $1.37 \pm 0.92$  & $0.81 \pm 0.33$ & $15.75 \pm 2.61$ & $58.28 \pm 11.99$ \\
        PtyNet   
        & $7.48 \pm 4.05$  & $1.52 \pm 0.71$ & $11.80 \pm 2.13$ & $58.07 \pm 15.77$
        & $3.52 \pm 1.62$  & $1.59 \pm 0.40$ & $11.36 \pm 2.04$ & $40.14 \pm 7.25$ \\
        PPN      
        & $4.85 \pm 3.19$  & $0.77 \pm 0.39$ & $14.28 \pm 3.45$ & $69.68 \pm 13.47$
        & $1.19 \pm 0.72$  & $0.78 \pm 0.30$ & $16.48 \pm 2.93$ & $50.15 \pm 8.96$ \\
        Ours
        & \textbf{\textcolor{forestgreen}{3.32 $\pm$ 2.51}}
        & \textbf{\textcolor{forestgreen}{0.40 $\pm$ 0.28}}
        & \textbf{\textcolor{forestgreen}{17.31 $\pm$ 3.37}}
        & \textbf{\textcolor{forestgreen}{85.69 $\pm$ 10.21}}
        & \textbf{\textcolor{forestgreen}{0.92 $\pm$ 0.59}}
        & \textbf{\textcolor{forestgreen}{0.70 $\pm$ 0.31}}
        & \textbf{\textcolor{forestgreen}{17.56 $\pm$ 2.18}}
        & \textbf{\textcolor{forestgreen}{73.91 $\pm$ 9.81}} \\
        \midrule
        \multicolumn{9}{l}{\textbf{Noisy}} \\
        \midrule
        PtychoNN
        & $17.39 \pm 12.25$ & $2.01 \pm 1.29$ & $8.83 \pm 3.48$ & $51.30 \pm 23.14$
        & $2.10 \pm 1.42$ & $1.04 \pm 0.45$ & $14.27 \pm 3.35$ & $48.14 \pm 19.90$ \\
        DPI
        & $7.63 \pm 4.44$ & $1.16 \pm 0.47$ & $11.92 \pm 2.64$ & $57.27 \pm 13.55$
        & $1.47 \pm 0.88$ & $0.84 \pm 0.33$ & $15.44 \pm 2.74$ & $56.60 \pm 12.03$ \\
        PtyNet
        & $9.30 \pm 4.59$ & $1.83 \pm 0.77$ & $10.82 \pm 2.14$ & $50.12 \pm 15.47$
        & $12.53 \pm 2.82$ & $3.28 \pm 0.45$ & $5.51 \pm 1.08$ & $28.51 \pm 8.35$ \\
        PPN
        & $5.09 \pm 2.96$ & $0.712 \pm 0.36$ & $14.26 \pm 3.17$ & $71.43 \pm 13.37$
        & $1.21 \pm 0.67$ & $0.80 \pm 0.29$ & $16.38 \pm 3.00$ & $48.58 \pm 7.38$ \\
        Ours
        & \textbf{\textcolor{forestgreen}{4.04 $\pm$ 2.96}}
        & \textbf{\textcolor{forestgreen}{0.47 $\pm$ 0.32}}
        & \textbf{\textcolor{forestgreen}{16.03 $\pm$ 3.21}}
        & \textbf{\textcolor{forestgreen}{82.94 $\pm$ 11.23}}
        & \textbf{\textcolor{forestgreen}{0.95 $\pm$ 0.55}}
        & \textbf{\textcolor{forestgreen}{0.73 $\pm$ 0.29}}
        & \textbf{\textcolor{forestgreen}{17.44 $\pm$ 2.80}}
        & \textbf{\textcolor{forestgreen}{73.10 $\pm$ 10.13}} \\
        \bottomrule
    \end{tabular}
    }
\end{table*}

\subsection{Per-Sample Reconstruction Quality}

\begin{figure}[!ht]
\centering
\includegraphics[width=1\linewidth]{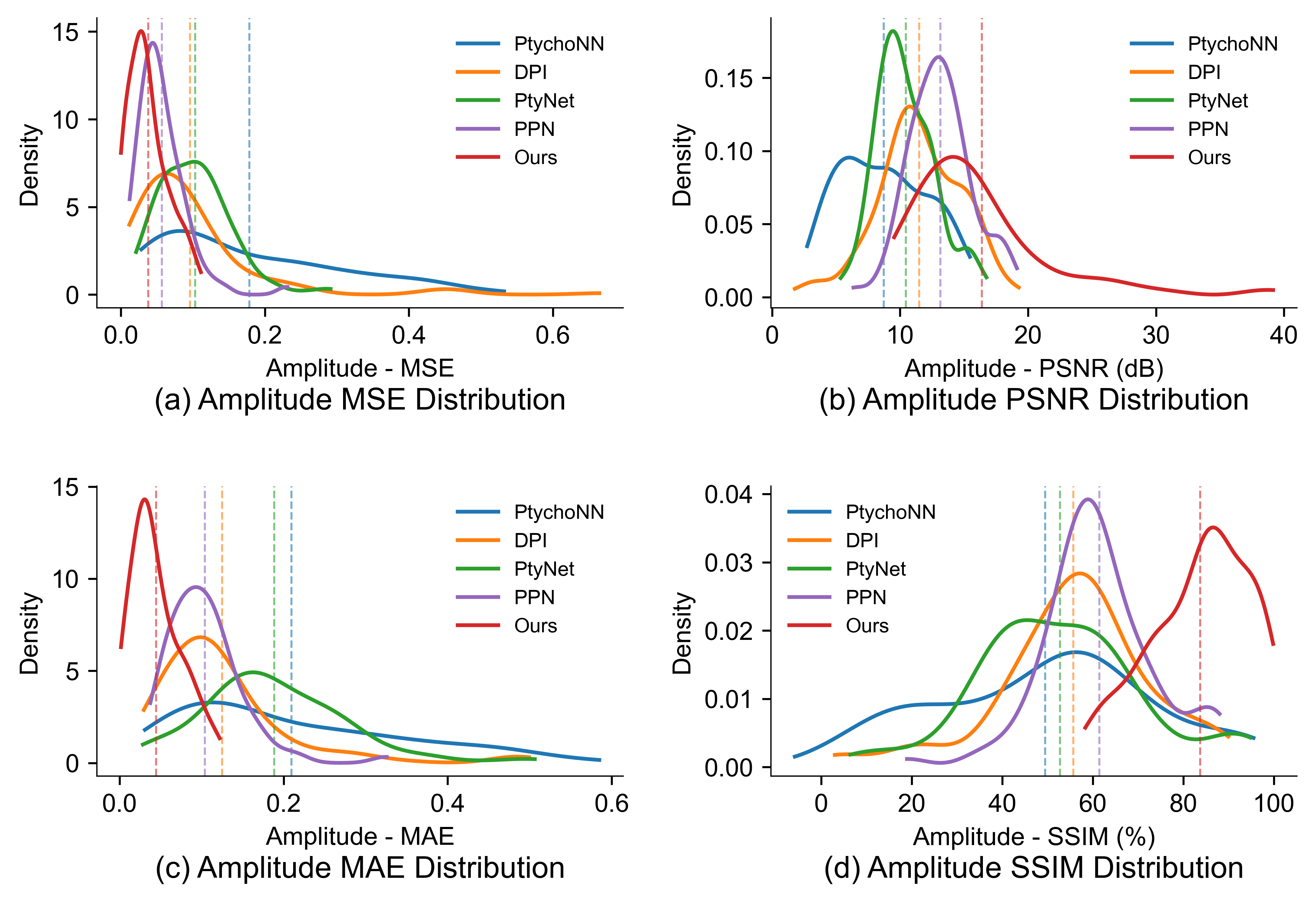}
\caption{Distribution of amplitude reconstruction metrics across methods on 144 test samples: (a) MSE, (b) PSNR, (c) MAE, and (d) SSIM. Vertical dashed lines mark the mean. Models: PtychoNN (blue), DPI (orange), PtyNet (green), PPN (purple), and Ours (red). Narrower distributions indicate more consistent performance; lower MSE/MAE and higher PSNR/SSIM indicate better reconstruction quality.}
\label{fig:distrub_amp} 
\end{figure}

\begin{figure}[!ht]
\centering
\includegraphics[width=1\linewidth]{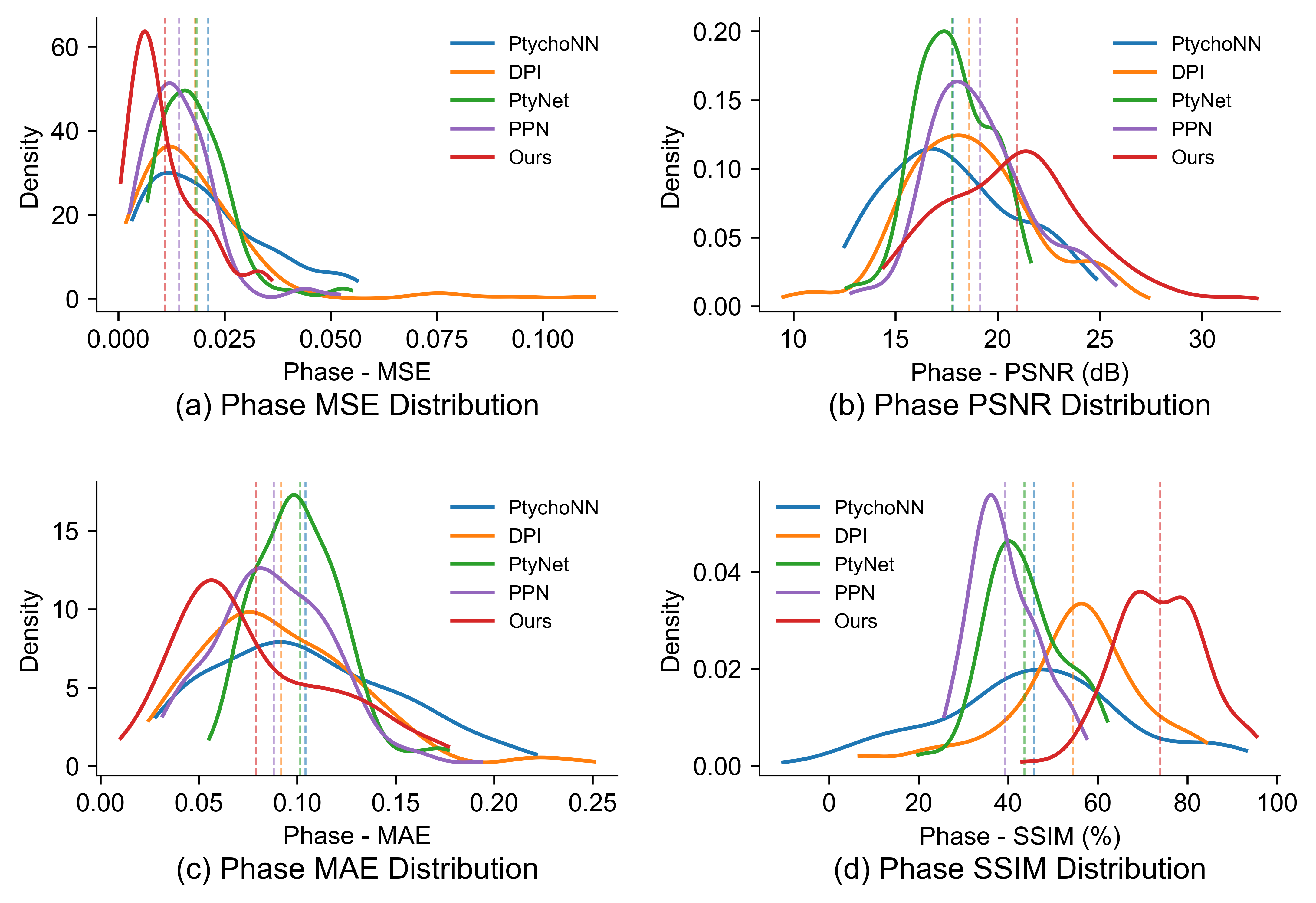}
\caption{Distribution of phase reconstruction metrics across methods on 144 test samples: (a) MSE, (b) PSNR, (c) MAE, and (d) SSIM. Vertical dashed lines mark the mean. Models: PtychoNN (blue), DPI (orange), PtyNet (green), PPN (purple), and Ours (red). Narrower distributions indicate more consistent performance; lower MSE/MAE and higher PSNR/SSIM indicate better reconstruction quality.}
\label{fig:distrub_ph} 
\end{figure}

Fig.~\ref{fig_single} shows single-shot amplitude and phase reconstructions for three representative test samples.
For amplitude in Fig.~\hyperref[fig_single]{2(a)}, the proposed method is closest to the ground truth, with sharper boundaries, cleaner piecewise-constant regions, and less leakage into the background.
PtychoNN and DPI tend to over-smooth high-contrast edges, leading to rounded corners and wider transition bands.
PtyNet shows clear structured artifacts and shape distortion, while PPN performs better than the earlier baselines but still exhibits local deformation and residual smoothing near bar edges.
Overall, our method preserves both geometry and contrast more faithfully.

The difference is more evident for phase in Fig.~\hyperref[fig_single]{2(b)}.
PtyNet often collapses to an over-smoothed, nearly constant phase map with artificial boundary framing, suggesting weak phase recovery in the single-shot setting.
PPN introduces visible texture-like noise and spatial inconsistency, while PtychoNN and DPI blur phase discontinuities and bias plateau levels.
By contrast, our method recovers cleaner and more coherent phase maps, with better aligned discontinuities and smoother backgrounds.

These visual trends are consistent with the metric distributions in Fig.~\ref{fig:distrub_amp} and Fig.~\ref{fig:distrub_ph}.
For amplitude, our method shifts the distributions toward lower MSE/MAE and higher PSNR/SSIM, while also reducing heavy-tail behavior compared with PtychoNN and DPI and remaining clearly separated from PtyNet.
Its PSNR distribution also extends further into the high-value region, indicating that more samples are reconstructed with very high fidelity.
For phase, our method similarly shifts PSNR and SSIM to the right and concentrates MSE at lower values, whereas DPI retains a noticeable long tail and PtyNet/PPN remain at clearly worse SSIM levels.

Table~\ref{tab:combined_metrics} summarizes per-sample performance (mean$\pm$std) under noise-free and noisy conditions.
In the noise-free setting, our method gives the best amplitude results, reaching PSNR $17.31$\,dB and SSIM $85.69\%$.
Compared with the strongest baseline, PPN, this corresponds to gains of $+3.03$\,dB PSNR and $+16.01$\,pp SSIM, together with MSE/MAE reductions of $31.5\%$ and $48.1\%$.
For phase, the largest advantage appears in structural quality: our SSIM reaches $73.91\%$, exceeding the best competing result (DPI, $58.28\%$) by $+15.63$\,pp, while reducing MSE from $1.19$ (PPN) to $0.92$.
Under noise, the proposed method remains the best performer for both amplitude and phase.
For amplitude, it improves on PPN by $+1.77$\,dB PSNR and $+11.51$\,pp SSIM, with a $20.6\%$ reduction in MSE.
For phase, it reaches SSIM $73.10\%$, outperforming the best baseline (DPI, $56.60\%$) by $+16.50$\,pp, while reducing MSE from $1.21$ (PPN) to $0.95$.

\begin{figure*}[!htbp]
\centering
\includegraphics[width=0.9\linewidth]{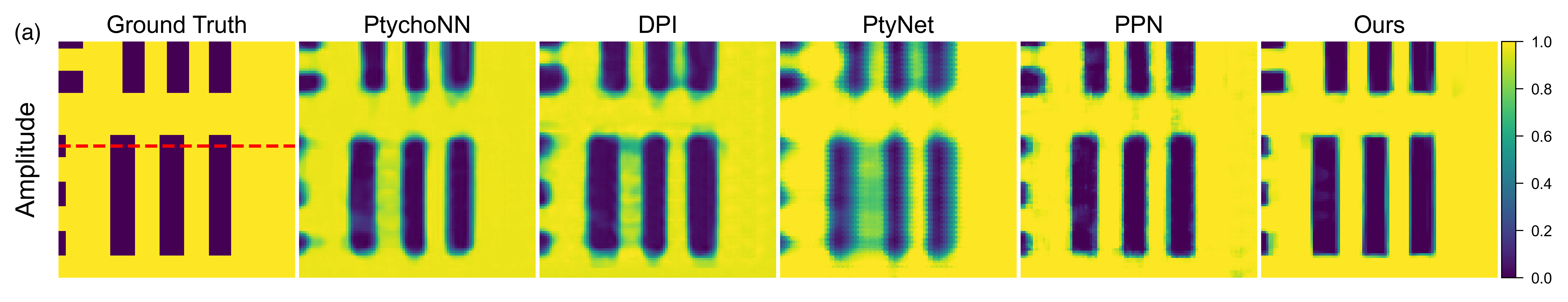}
\includegraphics[width=0.9\linewidth]{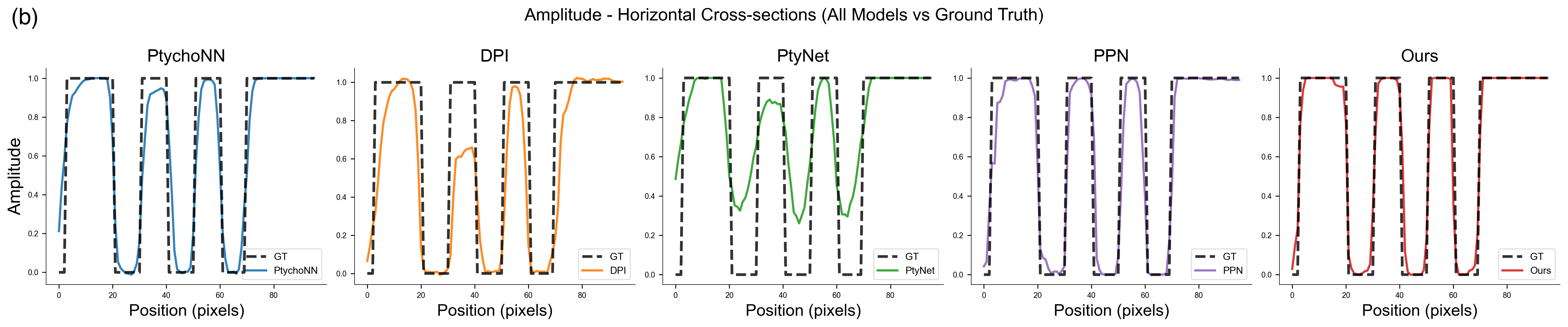}
\includegraphics[width=0.9\linewidth]{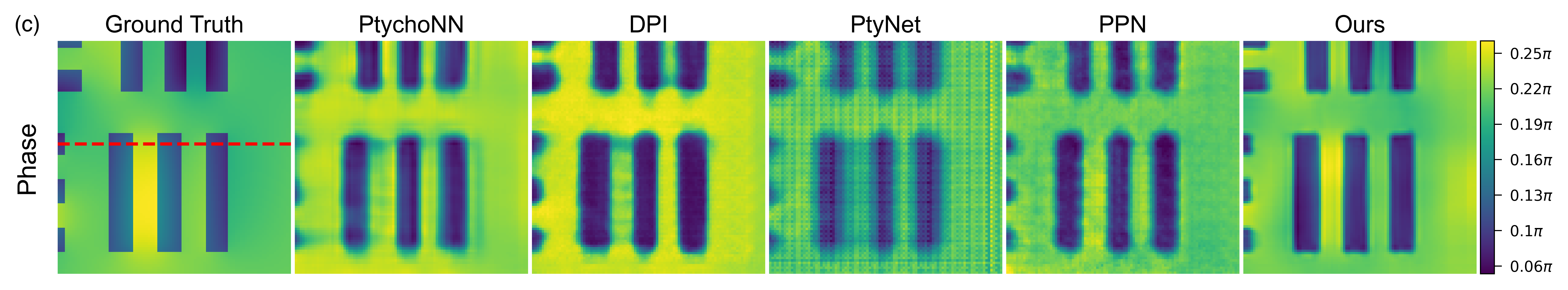}
\includegraphics[width=0.9\linewidth]{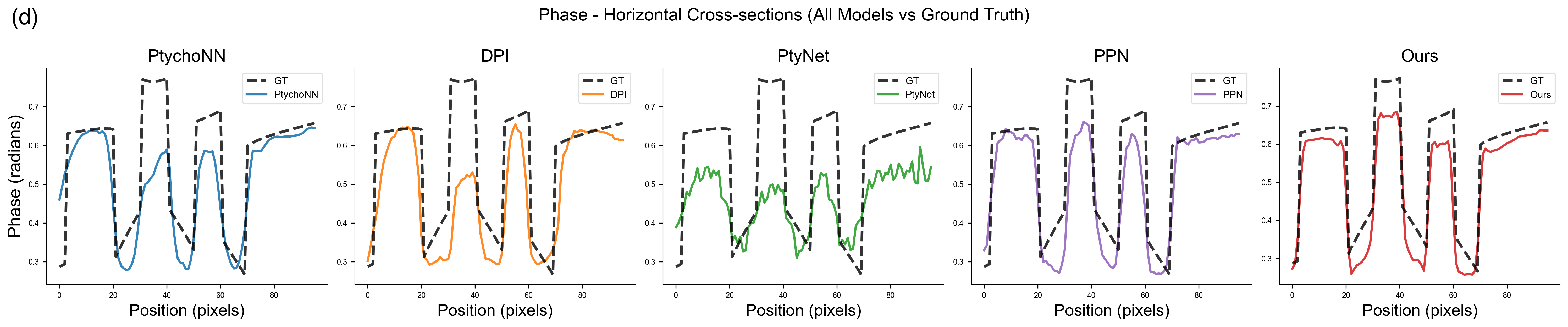}
\caption{Full-field amplitude and phase images stitched from single-shot reconstructions of different models. (a) Ground-truth and reconstructed amplitude maps from PtychoNN, DPI, PtyNet, PPN, and the proposed method; the red dashed line marks the horizontal cross-section used for profile extraction. (b) Amplitude profiles along that cross-section. (c) Ground-truth and reconstructed phase maps for all methods. (d) Phase profiles along the same cross-section.}
\label{fig_full}
\end{figure*}

\subsection{Stitched Full-Field Reconstruction}

To assess large-field-of-view reconstruction quality, the $12{\times}12$ grid of $32{\times}32$ patches is stitched into a composite image using an adaptive stitching algorithm.
A radially decaying weight kernel, $w(\mathbf{r})=(1-d(\mathbf{r})/d_{\max})^2$, is used within the overlap region defined by the scan step size (point size 8 pixels, overlap 40 pixels), which reduces seam artifacts and boundary discontinuities.

Stitched full-field evaluation places greater emphasis on global consistency and boundary handling, since local errors can accumulate during mosaicking.
Fig.~\ref{fig_full} compares stitched amplitude and phase maps together with horizontal cross-sections.
For amplitude (Fig.~\hyperref[fig_full]{5(a)--(b)}), our method yields sharper boundaries and more uniform backgrounds.
PtychoNN and DPI show rounded edges and reduced contrast, while PtyNet exhibits oscillatory artifacts.
PPN reduces some blur but still deviates from the ground-truth plateau levels near transitions.
The 1-D profiles show that our reconstruction follows the ground-truth step edges and plateau values more closely, with less overshoot and undershoot.

For phase (Fig.~\hyperref[fig_full]{5(c)--(d)}), the same trend holds.
PtyNet shows strong periodic grid artifacts, and PtychoNN/DPI introduce visible texture and biased plateau levels, which distort the cross-sectional profiles.
Our method produces the most coherent phase map and the closest agreement with the ground-truth profile, especially around discontinuities and flat regions, indicating better phase stability after stitching.

\begin{table*}[!h]
    \caption{Stitched-image reconstruction metrics for amplitude and phase. 
             MSE ($\times 10^{-2}$), MAE ($\times 10^{-1}$), PSNR (dB), SSIM (\%).}
    \label{tab:stitched_metrics}
    \centering
    \resizebox{0.8\textwidth}{!}{
    \begin{tabular}{@{\hspace{2mm}}l l cccccccc@{\hspace{2mm}}}
        \toprule
        \multirow{3}{*}{\textcolor{ac_gray}{\textbf{Conditions}}}
        & \multirow{3}{*}{\textcolor{ac_gray}{\textbf{Methods}}} 
        & \multicolumn{4}{c}{\textcolor{ac_gray}{\textbf{Amplitude}}} 
        & \multicolumn{4}{c}{\textcolor{ac_gray}{\textbf{Phase}}} \\
        \cmidrule(lr){3-6} \cmidrule(lr){7-10}
        & 
        & \textcolor{ac_gray}{MSE $\downarrow$} 
        & \textcolor{ac_gray}{MAE $\downarrow$} 
        & \textcolor{ac_gray}{PSNR $\uparrow$} 
        & \textcolor{ac_gray}{SSIM $\uparrow$}
        & \textcolor{ac_gray}{MSE $\downarrow$} 
        & \textcolor{ac_gray}{MAE $\downarrow$} 
        & \textcolor{ac_gray}{PSNR $\uparrow$} 
        & \textcolor{ac_gray}{SSIM $\uparrow$} \\ 
        &  & ($\times 10^{-2}$) & ($\times 10^{-1}$) & (dB) & (\%) & ($\times 10^{-2}$) & ($\times 10^{-1}$) & (dB) & (\%) \\
        \midrule
        \multirow{5}{*}{\textbf{Noise-Free}}
        & PtychoNN 
        & $3.90$ & $0.87$ & $14.09$ & $66.53$
        & $0.78$ & $0.61$ & $17.44$ & $67.66$ \\
        & DPI      
        & $3.29$ & $0.90$ & $14.83$ & $62.30$
        & $0.92$ & $0.69$ & $16.70$ & $66.32$ \\
        & PtyNet   
        & $4.83$ & $1.24$ & $13.16$ & $63.67$
        & $2.85$ & $1.48$ & $11.80$ & $50.46$ \\
        & PPN      
        & $2.36$ & $0.61$ & $16.27$ & $73.66$
        & $0.73$ & $0.61$ & $17.74$ & $65.70$ \\
        & Ours
        & \textbf{\textcolor{forestgreen}{1.29}} 
        & \textbf{\textcolor{forestgreen}{0.37}} 
        & \textbf{\textcolor{forestgreen}{18.88}} 
        & \textbf{\textcolor{forestgreen}{82.26}} 
        & \textbf{\textcolor{forestgreen}{0.40}} 
        & \textbf{\textcolor{forestgreen}{0.48}} 
        & \textbf{\textcolor{forestgreen}{20.33}} 
        & \textbf{\textcolor{forestgreen}{80.75}} \\
        \midrule
        \multirow{5}{*}{\textbf{Noisy}}
        & PtychoNN 
        & $7.97$ & $1.51$ & $10.99$ & $55.97$
        & $1.37$ & $0.83$ & $14.99$ & $58.72$ \\
        & DPI 
        & $3.43$ & $0.92$ & $14.64$ & $61.55$
        & $0.91$ & $0.69$ & $16.75$ & $65.21$ \\
        & PtyNet 
        & $6.41$ & $1.48$ & $11.93$ & $57.46$
        & $12.69$ & $3.37$ & $5.31$ & $38.04$ \\
        & PPN 
        & $2.35$ & $0.58$ & $16.30$ & $74.16$
        & $0.72$ & $0.61$ & $17.78$ & $63.58$ \\
        & Ours
        & \textbf{\textcolor{forestgreen}{1.47}}
        & \textbf{\textcolor{forestgreen}{0.38}}
        & \textbf{\textcolor{forestgreen}{18.33}}
        & \textbf{\textcolor{forestgreen}{81.85}}
        & \textbf{\textcolor{forestgreen}{0.49}}
        & \textbf{\textcolor{forestgreen}{0.57}}
        & \textbf{\textcolor{forestgreen}{19.42}}
        & \textbf{\textcolor{forestgreen}{80.02}} \\
        \bottomrule
    \end{tabular}
    }
\end{table*}

Table~\ref{tab:stitched_metrics} reports stitched full-field metrics, which reflect global reconstruction quality after mosaicking.
In the noise-free setting, our method improves on PPN by $+2.61$\,dB and $+8.60$\,pp for amplitude, with a $45.3\%$ reduction in MSE.
For phase, it reaches SSIM $80.75\%$, exceeding the best competing SSIM (PtychoNN, $67.66\%$) by $+13.09$\,pp, while reducing MSE by $45.2\%$ relative to PPN.
Under noise, the advantage remains.
For amplitude, our method improves on PPN by $+2.03$\,dB and $+7.69$\,pp, with a $37.4\%$ reduction in MSE.
For phase, it reaches SSIM $80.02\%$, outperforming the best baseline (DPI, $65.21\%$) by $+14.81$\,pp.

\begin{figure*}[!htbp]
\centering
\includegraphics[width=0.9\linewidth]{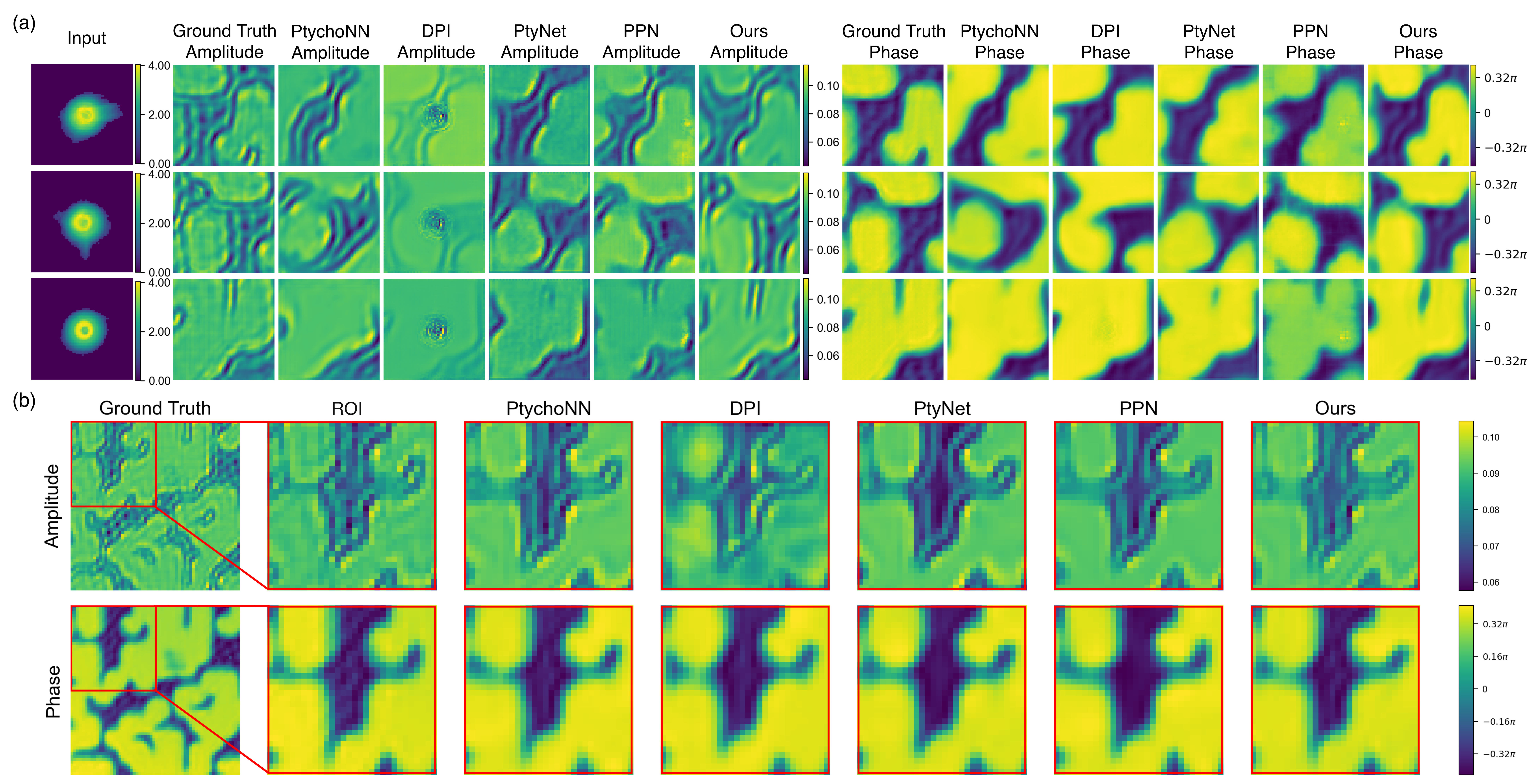}
\caption{Qualitative comparison of amplitude and phase reconstructions.
(a) Reconstruction results for representative test samples. From left to right are the input diffraction intensity, the ground-truth amplitude, and the amplitude reconstructed by PtychoNN, DPI, PtyNet, PPN, and the proposed method (Ours). The corresponding phase reconstructions are shown on the right.
(b) Comparison on stitched reconstructions used to visualize mid-frequency content and image texture quality over a larger field of view. The stitched ground truth is shown in the leftmost column, with regions of interest (ROIs) highlighted by red boxes. Zoomed-in views of the stitched amplitude (top row) and phase (bottom row) are shown for each method. The proposed method exhibits clearer mid-frequency structures and more consistent texture patterns that are visually closer to the ground truth.}
\label{fig:real}
\end{figure*}

\subsection{Validation on Experimental Synchrotron Data}

To assess robustness to real detector/ex\-perimental effects and generalization to a different diffraction input resolution ($64{\times}64$), we further validate the competing methods on experimental synchrotron data.

Fig.~\ref{fig:real} shows qualitative comparisons in both the single-shot and stitched settings.
In Fig.~\ref{fig:real}(a), several baselines show clear limitations on real measurements.
DPI leaves strong residual artifacts and washed-out amplitude contrast, with fine streak-like structures largely suppressed.
PtychoNN and PPN recover the main shapes but over-smooth mid-scale ridges, broadening thin structures and weakening local contrast.
PtyNet preserves some edges but introduces blocky textures inconsistent with the ground truth.
By contrast, our method better preserves elongated filamentary patterns in amplitude and produces cleaner, more consistent phase maps with sharper boundaries and fewer spurious fluctuations.

Fig.~\ref{fig:real}(b) compares stitched reconstructions over a larger FOV, with the ROI marked in red.
In the zoomed region, our amplitude reconstruction retains thin branching structures and curvature patterns that are blurred in PtychoNN/PPN, strongly attenuated in DPI, or distorted by block artifacts in PtyNet.
For phase, the difference is more apparent: PtyNet and PPN show deep-black texture-like patterns over large regions, suggesting corruption of mid-frequency phase content.
Our method preserves more coherent phase structures, smoother background variation, and more accurate transition locations, remaining visually closest to the ground truth.

Table~\ref{tab:stitched_metrics_real} confirms the same trend on stitched experimental reconstructions.
For amplitude, our method achieves the best MSE and MAE, and improves PSNR/SSIM to $24.32$\,dB/$83.92\%$, exceeding the strongest baseline by $+0.87$\,dB and $+3.60$\,pp, with a $13.3\%$ reduction in MSE.
The gains are larger for phase: MSE drops from $3.26$ to $2.07$ ($36.5\%$ reduction), while PSNR/SSIM increase to $26.14$\,dB/$90.54\%$, improving on the best baseline by $+1.97$\,dB and $+3.00$\,pp.
Overall, these results show good generalization to experimental data, especially for phase fidelity and mid-frequency structure preservation.

\begin{table}[!ht]
\caption{Stitched-image reconstruction metrics for amplitude and phase.
MSE and MAE are reported in absolute values, PSNR in dB, and SSIM in \%.}
\label{tab:stitched_metrics_real}
\centering
\setlength{\tabcolsep}{4pt}
\resizebox{0.4\textwidth}{!}{
\begin{tabular}{l ccccc}
\toprule
\textcolor{ac_gray}{\textbf{Metrics}} 
& \textcolor{ac_gray}{\textbf{PtychoNN}}
& \textcolor{ac_gray}{\textbf{DPI}}
& \textcolor{ac_gray}{\textbf{PtyNet}}
& \textcolor{ac_gray}{\textbf{PPN}}
& \textcolor{ac_gray}{\textbf{Ours}} \\
\midrule
\multicolumn{6}{l}{\textbf{Amplitude}} \\
\midrule
MSE($\times 10^{-4}$) $\downarrow$
& 0.15
& 6.27
& 0.19
& 0.16
& \textbf{\textcolor{forestgreen}{0.13}} \\
MAE($\times 10^{-2}$) $\downarrow$
& 0.27
& 2.42
& 0.30
& 0.27
& \textbf{\textcolor{forestgreen}{0.24}} \\
PSNR(dB) $\uparrow$
& 23.45
& 10.38
& 22.56
& 23.39
& \textbf{\textcolor{forestgreen}{24.32}} \\
SSIM(\%) $\uparrow$
& 80.32
& 46.31
& 74.66
& 80.10
& \textbf{\textcolor{forestgreen}{83.92}} \\
\midrule
\multicolumn{6}{l}{\textbf{Phase}} \\
\midrule
MSE($\times 10^{-2}$) $\downarrow$
& 3.54
& 7.20
& 3.98
& 3.26
& \textbf{\textcolor{forestgreen}{2.07}} \\
MAE($\times 10^{-1}$) $\downarrow$
& 1.48
& 2.20
& 1.59
& 1.40
& \textbf{\textcolor{forestgreen}{1.14}} \\
PSNR(dB) $\uparrow$
& 23.81
& 20.73
& 23.30
& 24.17
& \textbf{\textcolor{forestgreen}{26.14}} \\
SSIM(\%) $\uparrow$
& 86.72
& 75.66
& 85.34
& 87.54
& \textbf{\textcolor{forestgreen}{90.54}} \\
\bottomrule
\end{tabular}
}
\end{table}

\begin{figure}[!h]
\centering
\includegraphics[width=0.493\linewidth]{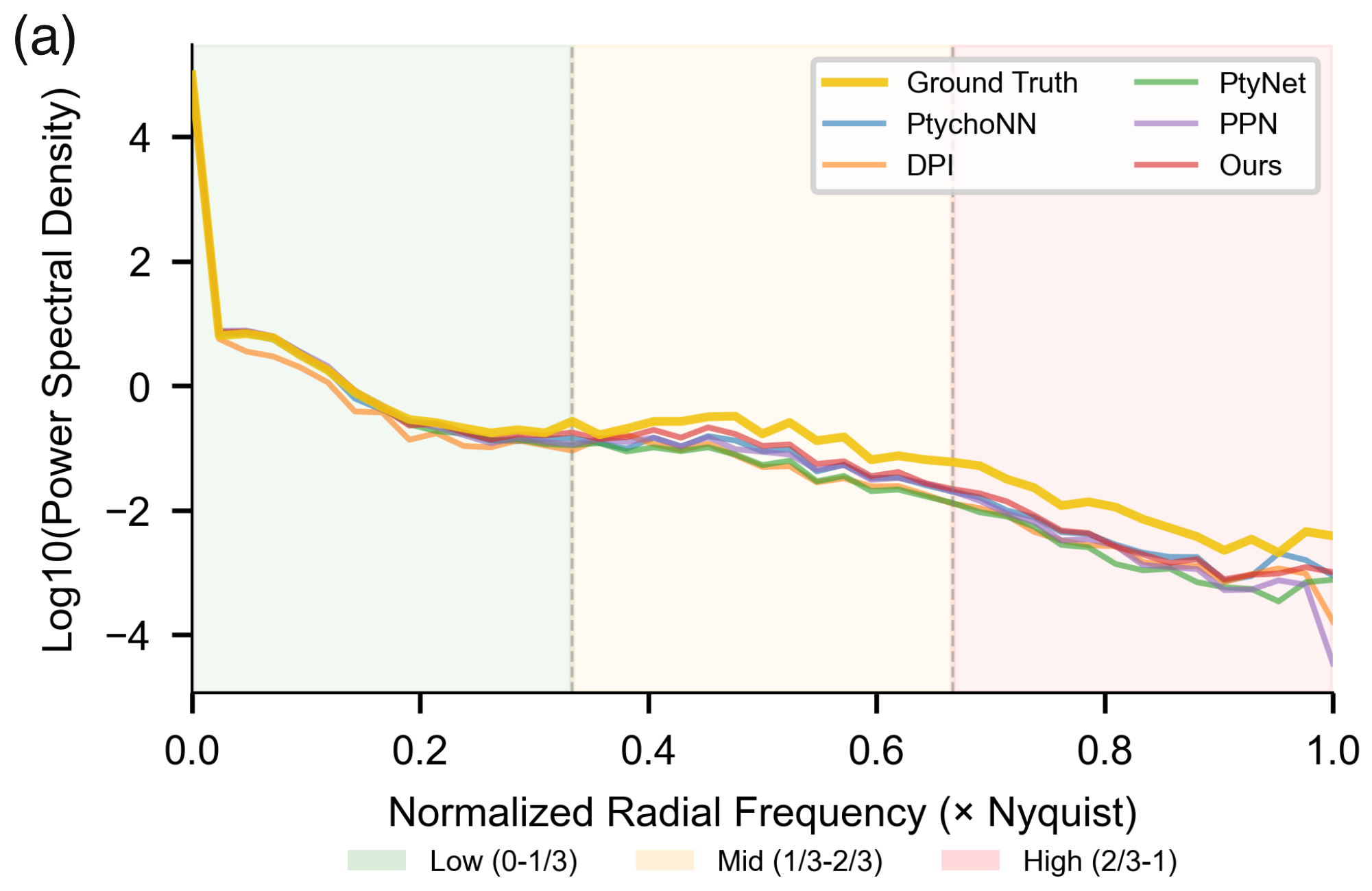}
\includegraphics[width=0.493\linewidth]{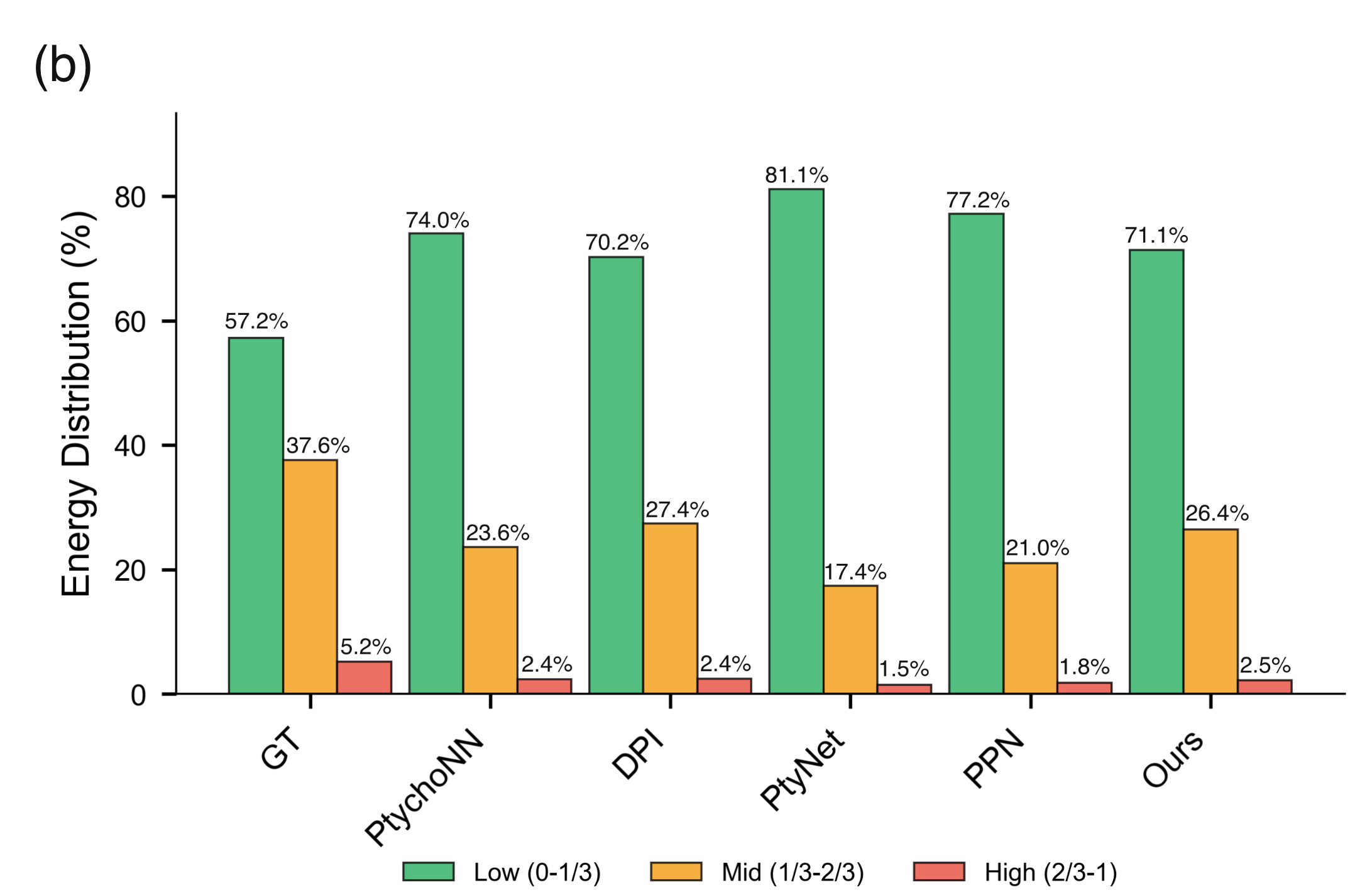}
\caption{Amplitude spectral analysis.
(a) Radially averaged power spectral density (PSD, log scale) of reconstructed amplitude images from different methods as a function of normalized spatial frequency. The ground truth is shown for reference, and shaded regions indicate low-, mid-, and high-frequency bands.
(b) Energy distribution (\%) of amplitude reconstructions across frequency bands.}
\label{fig:spatial_amplitude}
\end{figure}

\subsection{Spatial Frequency Analysis}

To complement pixel-wise metrics, we analyze spatial-frequency fidelity on the stitched full-field reconstructions.
For each result, we compute the 2-D power spectral density (PSD) using the FFT and extract a 1-D radially averaged PSD.
Following Fig.~\ref{fig:spatial_amplitude} and Fig.~\ref{fig:spatial_phase}, the spectral energy is divided into three bands: low ($0$--$\frac{1}{3}$ Nyquist), mid ($\frac{1}{3}$--$\frac{2}{3}$ Nyquist), and high ($\frac{2}{3}$--$1$ Nyquist), with each radial bin weighted by its radius to account for the increasing number of Fourier coefficients at larger frequencies.

For amplitude, Fig.~\ref{fig:spatial_amplitude}(a) shows that our radially averaged PSD follows the ground-truth decay more closely in the mid and high bands, whereas the baselines fall below the ground truth.
This trend is also reflected in Fig.~\ref{fig:spatial_amplitude}(b): the ground truth contains $37.6\%$ mid-band and $5.2\%$ high-band energy, while most baselines shift more energy to the low band.
For example, PtyNet allocates $81.1\%$ to the low band, $17.4\%$ to the mid band, and only $1.5\%$ to the high band.
Our method retains more mid/high-frequency energy (mid $26.4\%$, high $2.5\%$), improving over PPN by $+5.4$\,pp in the mid band and $+0.7$\,pp in the high band, consistent with the sharper edges and clearer textures in the stitched reconstructions.

\begin{figure}[!htbp]
\centering
\includegraphics[width=0.493\linewidth]{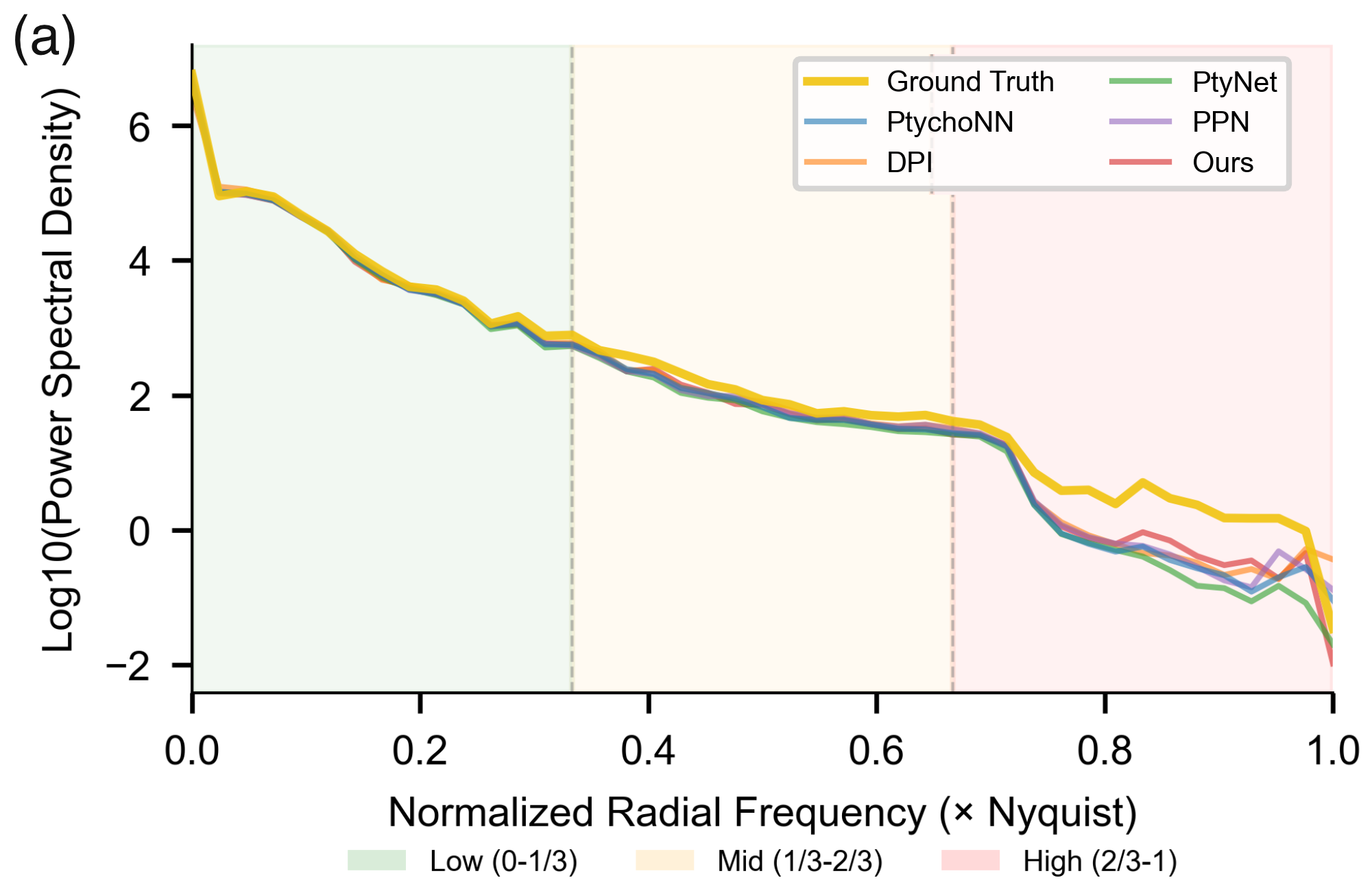}
\includegraphics[width=0.493\linewidth]{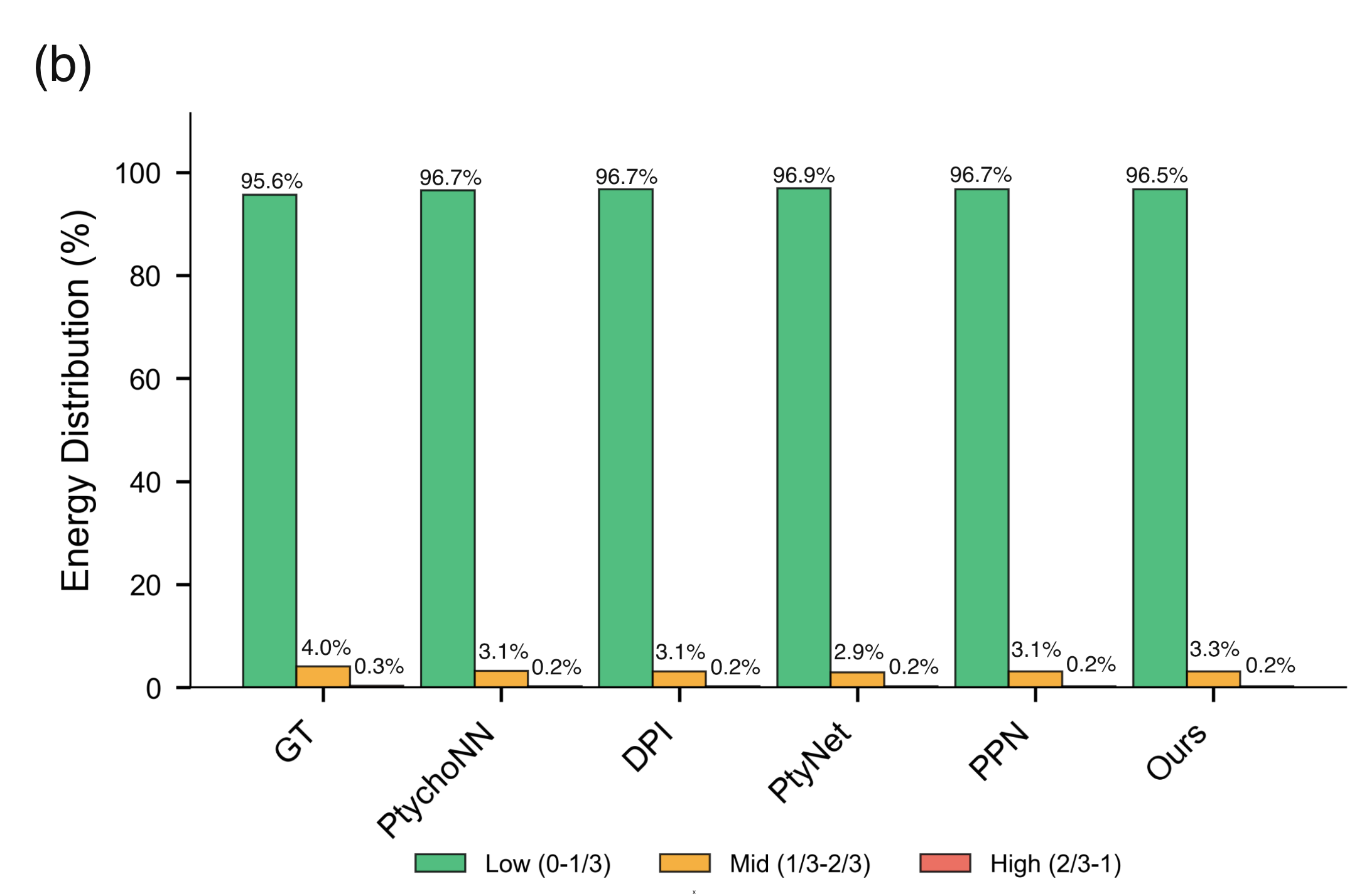}
\caption{Phase spectral analysis.
(a) Radially averaged power spectral density (PSD, log scale) of reconstructed phase images from different methods as a function of normalized spatial frequency. The ground truth is shown for reference, and shaded regions indicate low-, mid-, and high-frequency bands.
(b) Energy distribution $(\%)$ of phase reconstructions across frequency bands.}
\label{fig:spatial_phase}
\end{figure}

For phase, Fig.~\ref{fig:spatial_phase}(a) shows that all methods are similar at low frequencies, reflecting the smooth overall phase structure, but clearer differences appear beyond $\frac{2}{3}$ Nyquist, where baseline spectra decay more rapidly.
Fig.~\ref{fig:spatial_phase}(b) shows that the ground truth contains $4.0\%$ mid-band and $0.3\%$ high-band energy.
All reconstructions reduce these components, but our method gives the closest mid-band allocation among the learned models (mid $3.3\%$) while keeping the high-band energy low ($0.2\%$), suggesting better recovery of meaningful mid-frequency phase variation without introducing excessive high-frequency noise.

Overall, Fig.~\ref{fig:spatial_amplitude} and Fig.~\ref{fig:spatial_phase} show that the proposed method preserves informative mid-to-high frequency content more effectively, especially for amplitude and mid-band phase structure, consistent with the visual sharpness and the loss design.

\begin{figure*}[!htbp]
\centering
\includegraphics[width=0.9\linewidth]{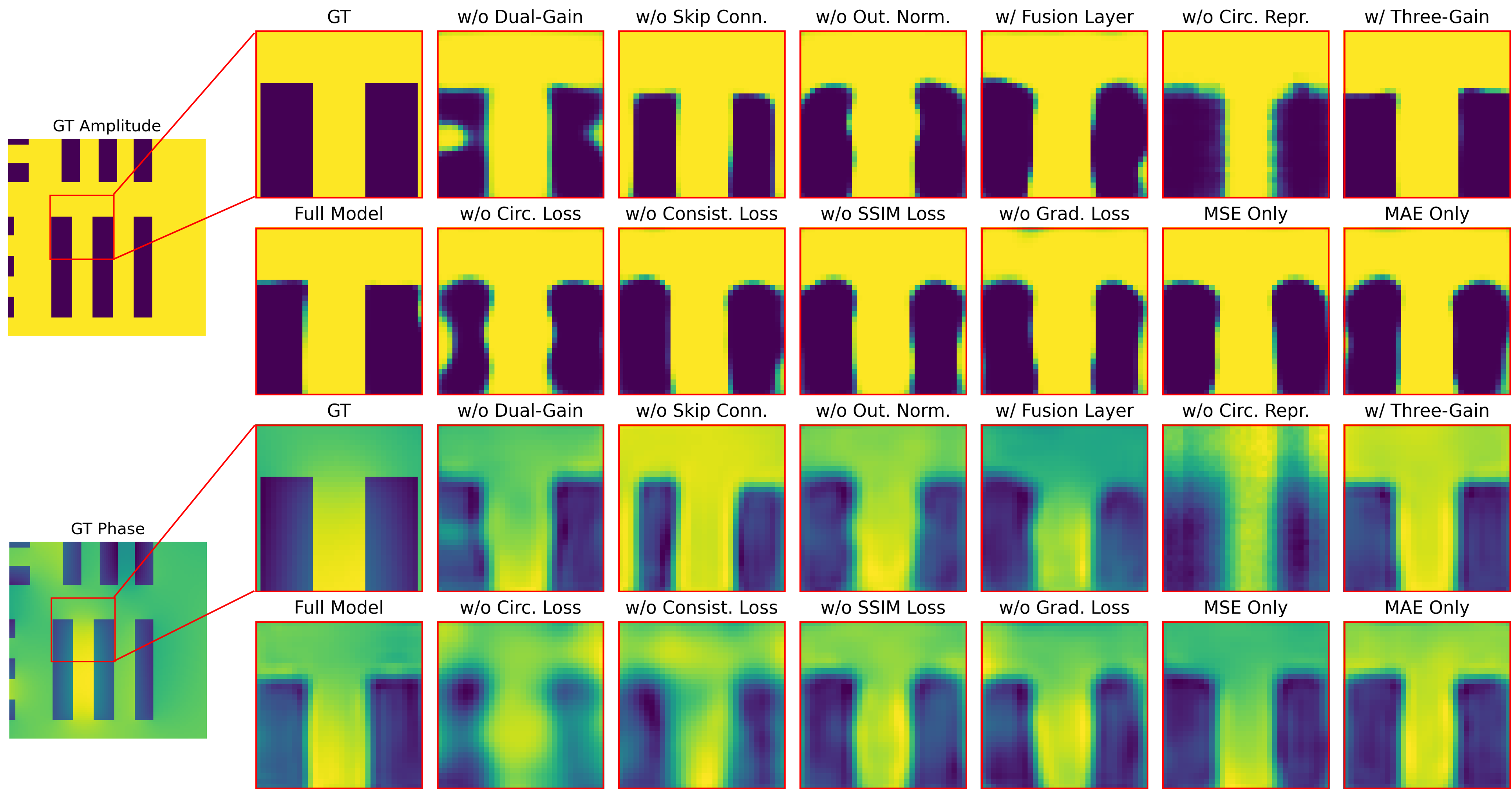}
\caption{Visual Comparison of Amplitude and Phase Reconstructions --- Architectural and Loss Function Ablation Studies with ROI (region of interest) Zoom. 1st row: Amplitude (Architectural Ablation); 2nd row: Amplitude (Loss Function Ablation); 3rd row: Phase (Architectural Ablation); 4th row: Phase (Loss Function Ablation).}
\label{fig:ablation_roi}
\end{figure*}

\section{Ablation Study}
\label{sec:ablation}

A systematic ablation study is conducted to quantify the contribution of individual architectural components and loss terms to the final reconstruction quality.
All variants are trained and evaluated under the same protocol as the full model, which serves as the reference baseline.
The study is organized into two groups: architectural ablation (modifying network components) and loss function ablation (modifying the training objective).

\begin{figure}[!ht]
    \centering
    \includegraphics[width=0.9\linewidth]{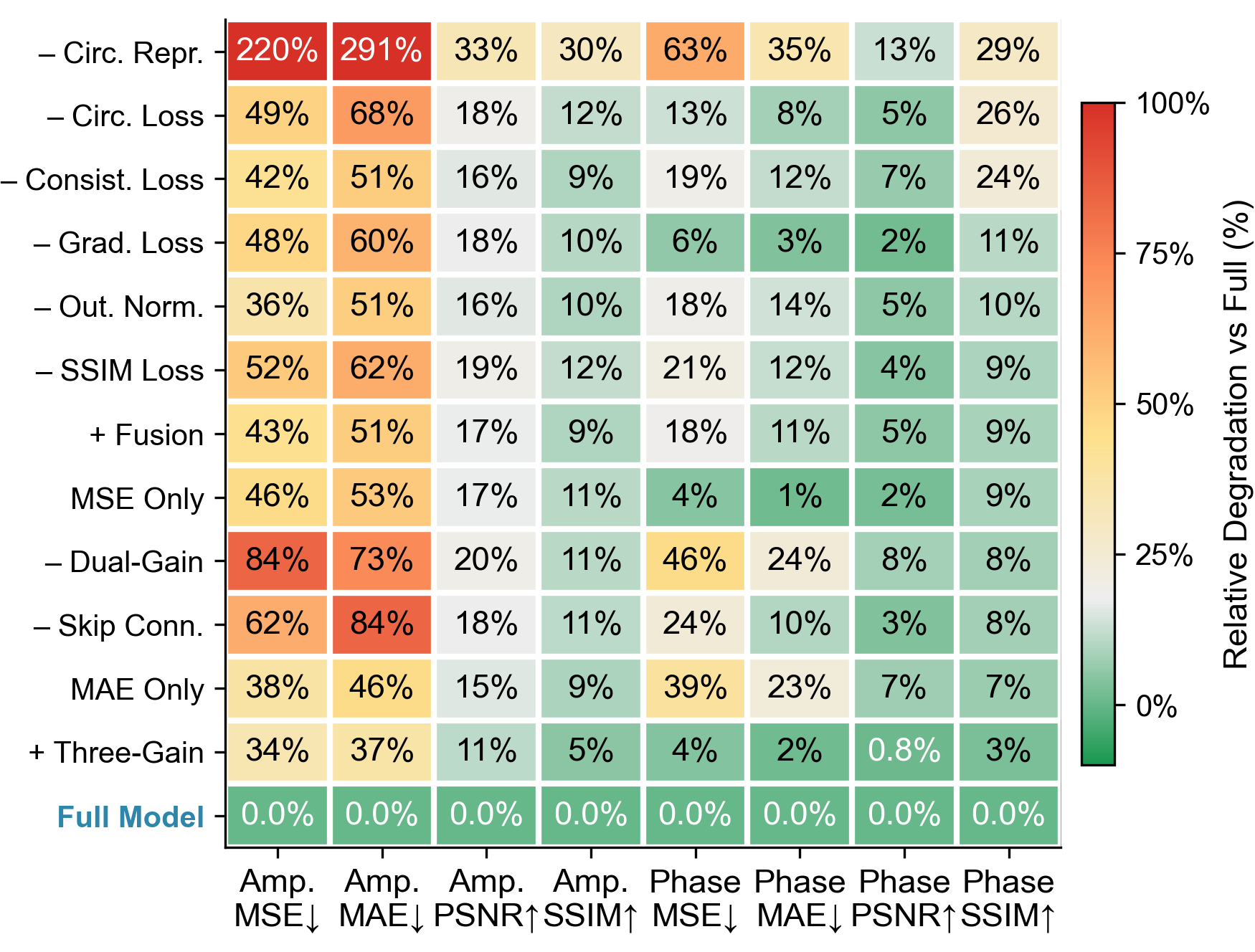}
    \caption{Ablation analysis of model components. Heatmap showing the relative degradation $(\%)$ in amplitude and phase reconstruction metrics with respect to the Full Model. Each row corresponds to the removal or modification of a specific loss term or architectural component, and columns report changes in MSE, MAE, PSNR, and SSIM for amplitude and phase.}
    \label{fig:heatmap}
\end{figure}

\subsection{Architectural Ablation}
\label{sec:arch_ablation}

We evaluate six architectural variants.
1) \emph{w/o Dual-Gain}: removes the SADGS module and uses a single-gain input, so strong and weak diffraction features can no longer be emphasized simultaneously.
2) \emph{w/o Skip Conn.}: removes the shared learnable skip connection, forcing the decoders to recover spatial detail only from the bottleneck representation.
3) \emph{w/o Out.\ Norm.}: removes the unit-circle normalization in \eqref{eq:unit_norm}, leaving the cosine and sine outputs $\tanh$-bounded but not explicitly projected onto $S^1$.
4) \emph{w/ Fusion Layer}: replaces the $1{\times}1$ projection with a deeper convolutional fusion module to test whether extra fusion capacity is beneficial.
5) \emph{w/o Circ.\ Repr.}: removes the circular phase representation and instead predicts phase directly as a scalar with a single decoder and a $\tanh$ output scaled to $(-\pi,\pi]$, trained with Euclidean MSE loss.
6) \emph{w/ Three-Gain}: extends the dual-gain module to three gain levels to test whether an additional dynamic-range view improves performance.

\subsection{Loss Function Ablation}
\label{sec:loss_ablation}

We evaluate six loss variants.
1) \emph{w/o Circ.\ Loss}: removes the geodesic circular loss by setting $\lambda_{\mathrm{circ}}=0$, so phase is optimized only with component-wise Euclidean losses.
2) \emph{w/o Consist.\ Loss}: removes the manifold consistency term by setting $w_c=0$.
3) \emph{w/o SSIM Loss}: removes the SSIM term by setting $\lambda_s=0$ for both amplitude and phase.
4) \emph{w/o Grad.\ Loss}: removes the gradient preservation term by setting $\lambda_g=0$.
5) \emph{MSE Only}: removes all auxiliary losses and trains using only MSE on all output channels.
6) \emph{MAE Only}: same as above, but using MAE on all output channels.

\subsection{Ablation Results and Analysis}

Fig.~\ref{fig:ablation_roi} shows qualitative comparisons of the ablation variants, and Table~\ref{tab:ablation-all} reports the corresponding quantitative results (mean$\pm$std).
For a clearer comparison, Fig.~\ref{fig:heatmap} summarizes the relative degradation of each variant with respect to the Full Model.

\subsubsection{Circular phase representation is the dominant factor}

Removing the circular phase representation (\emph{w/o Circ.\ Repr.}) causes the largest overall degradation.
As shown in Fig.~\ref{fig:ablation_roi}, this variant produces heavily corrupted phase maps with strong streaking and texture artifacts, together with blurred amplitude boundaries.
Quantitatively, amplitude MSE increases from $3.32$ to $10.61$ ($+219.6\%$), amplitude SSIM drops from $85.69\%$ to $59.63\%$, and phase SSIM drops from $71.80\%$ to $51.04\%$.
It also shows substantially larger variance, for example an amplitude SSIM std of $24.26\%$ versus $10.21\%$ for the Full Model.
This trend is also clear in Fig.~\ref{fig:heatmap}, confirming that circular phase modeling is central to the method rather than an auxiliary design choice.

\subsubsection{Phase-geometry losses are critical}

Among the loss ablations, removing the geodesic circular loss (\emph{w/o Circ.\ Loss}) and the manifold consistency loss (\emph{w/o Consist.\ Loss}) most strongly degrades phase structure.
Both variants show weaker phase contrast and less coherent transitions than the Full Model.
In Table~\ref{tab:ablation-all}, phase SSIM drops to $53.20\%$ and $54.74\%$, corresponding to relative degradations of about $26\%$ and $24\%$.
These results indicate that both explicit circular-distance supervision and manifold consistency are important for structurally correct phase reconstruction.

\subsubsection{Key architectural components: dual-gain and skip connection}

Among the architectural ablations, removing dual-gain scaling (\emph{w/o Dual-Gain}) causes the largest degradation.
It produces softer amplitude transitions and less stable phase structure, with relative degradations of $+84\%/+73\%$ in amplitude MSE/MAE and $+46\%$ in phase MSE.
Removing the skip connection (\emph{w/o Skip Conn.}) also reduces reconstruction sharpness, increasing amplitude MSE/MAE by $+62\%/+84\%$, which shows that direct input-to-decoder information flow is important for fine detail recovery.
Removing output normalization (\emph{w/o Out.\ Norm.}) mainly affects phase quality, with phase MAE increasing by $14\%$ and phase SSIM dropping by $10\%$, consistent with the role of explicit projection onto $S^1$.

Replacing the simple $1{\times}1$ projection with a deeper fusion module (\emph{w/ Fusion Layer}) does not help and slightly worsens most metrics, suggesting that extra fusion capacity is unnecessary in this setting.
Extending dual-gain to three gains (\emph{w/ Three-Gain}) yields only marginal changes and remains close to the Full Model, indicating that two gain levels are already sufficient.

\subsubsection{Role of perceptual and edge-aware supervision}

Removing SSIM loss (\emph{w/o SSIM Loss}) or gradient loss (\emph{w/o Grad.\ Loss}) mainly affects amplitude sharpness and texture fidelity.
This is reflected by the large amplitude degradations in Fig.~\ref{fig:heatmap} and by the visibly smoother boundaries in Fig.~\ref{fig:ablation_roi}.
Single-term objectives are also consistently weaker: both \emph{MSE Only} and \emph{MAE Only} underperform the Full Model across amplitude and phase metrics, supporting the use of the composite objective.

\subsubsection{Summary}

Overall, the ablation results show a clear importance ordering.
The circular phase representation is the most important component overall; the circular geodesic and consistency losses are the key loss terms for phase quality; dual-gain scaling and the skip connection are the most important architectural components for preserving detail; and SSIM and gradient supervision provide additional gains in sharpness and texture fidelity.
By contrast, deeper fusion and an extra gain branch offer little benefit.

\begin{table*}[!ht]
\centering
\caption{Ablation Study Results. MSE scaled by $\times 10^{-2}$, MAE scaled by $\times 10^{-1}$, SSIM as percentage. Values shown as mean $\pm$ std. \textcolor{forestgreen}{Best} and \textcolor{forestred}{worst} results highlighted. $\uparrow$ higher is better, $\downarrow$ lower is better.}
\label{tab:ablation-all}
\begingroup
\small
\setlength{\tabcolsep}{3pt}
\resizebox{0.8\textwidth}{!}{%
\begin{tabular}{@{}lcccccccc@{}}
\toprule
\textbf{Experiments} & \multicolumn{4}{c}{\textbf{Amplitude}} & \multicolumn{4}{c}{\textbf{Phase}} \\
\cmidrule(lr){2-5} \cmidrule(lr){6-9}
& MSE$\downarrow$ & MAE$\downarrow$ & PSNR$\uparrow$ & SSIM$\uparrow$ & MSE$\downarrow$ & MAE$\downarrow$ & PSNR$\uparrow$ & SSIM$\uparrow$ \\
& ($\times 10^{-2}$) & ($\times 10^{-1}$) & (dB) & (\%) & ($\times 10^{-2}$) & ($\times 10^{-1}$) & (dB) & (\%) \\
\midrule
w/o Dual-Gain 
& 6.10 $\pm$ 4.04 & 0.69 $\pm$ 0.43 & 13.91 $\pm$ 5.63 & 76.30 $\pm$ 13.12 
& 1.35 $\pm$ 0.92 & 0.86 $\pm$ 0.38 & 16.14 $\pm$ 3.30 & 66.19 $\pm$ 13.31 \\
w/o Skip Conn. 
& 5.37 $\pm$ 3.78 & 0.73 $\pm$ 0.46 & 14.28 $\pm$ 4.65 & 76.18 $\pm$ 13.22 
& 1.14 $\pm$ 0.83 & 0.77 $\pm$ 0.35 & 16.97 $\pm$ 3.49 & 66.21 $\pm$ 12.28 \\
w/o Out. Norm. 
& 4.51 $\pm$ 2.89 & 0.60 $\pm$ 0.34 & 14.51 $\pm$ 3.70 & 77.40 $\pm$ 11.59 
& 1.09 $\pm$ 0.52 & 0.80 $\pm$ 0.28 & 16.62 $\pm$ 2.59 & 64.38 $\pm$ 12.04 \\
w/ Fusion Layer 
& 4.76 $\pm$ 2.78 & 0.60 $\pm$ 0.32 & 14.28 $\pm$ 3.61 & 77.57 $\pm$ 10.58 
& 1.09 $\pm$ 0.59 & 0.77 $\pm$ 0.28 & 16.66 $\pm$ 2.52 & 65.49 $\pm$ 9.54 \\
w/o Circ. Repr. 
& \textcolor{red}{10.61 $\pm$ 8.61} & \textcolor{red}{1.56 $\pm$ 1.21} & \textcolor{red}{11.59 $\pm$ 7.15} & \textcolor{red}{59.63 $\pm$ 24.26} 
& \textcolor{red}{1.50 $\pm$ 0.90} & \textcolor{red}{0.94 $\pm$ 0.34} & \textcolor{red}{15.31 $\pm$ 2.53} & \textcolor{red}{51.04 $\pm$ 14.85} \\
w/ Three-Gain 
& 4.44 $\pm$ 3.38 & 0.55 $\pm$ 0.39 & 15.34 $\pm$ 4.90 & 81.48 $\pm$ 12.34 
& 0.96 $\pm$ 0.57 & 0.71 $\pm$ 0.26 & 17.42 $\pm$ 3.04 & 69.87 $\pm$ 11.44 \\
\midrule
w/o Circ. Loss 
& 4.96 $\pm$ 2.73 & 0.67 $\pm$ 0.33 & 14.11 $\pm$ 3.78 & 75.24 $\pm$ 11.33 
& 1.04 $\pm$ 0.48 & 0.76 $\pm$ 0.20 & 16.67 $\pm$ 3.08 & 53.20 $\pm$ 8.82 \\
w/o Consist. Loss 
& 4.72 $\pm$ 2.89 & 0.60 $\pm$ 0.33 & 14.55 $\pm$ 4.20 & 77.66 $\pm$ 11.24 
& 1.10 $\pm$ 0.51 & 0.78 $\pm$ 0.22 & 16.40 $\pm$ 2.05 & 54.74 $\pm$ 9.54 \\
w/o SSIM Loss 
& 5.05 $\pm$ 2.81 & 0.65 $\pm$ 0.33 & 14.04 $\pm$ 3.79 & 75.43 $\pm$ 12.22 
& 1.11 $\pm$ 0.61 & 0.78 $\pm$ 0.31 & 16.79 $\pm$ 3.16 & 65.19 $\pm$ 10.79 \\
w/o Grad. Loss 
& 4.90 $\pm$ 2.90 & 0.64 $\pm$ 0.34 & 14.24 $\pm$ 3.85 & 76.83 $\pm$ 11.79 
& 0.97 $\pm$ 0.51 & 0.72 $\pm$ 0.24 & 17.23 $\pm$ 2.85 & 64.05 $\pm$ 11.00 \\
MSE Only 
& 4.84 $\pm$ 2.98 & 0.61 $\pm$ 0.33 & 14.35 $\pm$ 3.98 & 76.57 $\pm$ 11.91 
& 0.96 $\pm$ 0.48 & 0.71 $\pm$ 0.23 & 17.15 $\pm$ 2.50 & 65.62 $\pm$ 10.39 \\
MAE Only 
& 4.57 $\pm$ 2.80 & 0.58 $\pm$ 0.32 & 14.63 $\pm$ 4.08 & 77.89 $\pm$ 10.81 
& 1.29 $\pm$ 0.99 & 0.86 $\pm$ 0.38 & 16.26 $\pm$ 2.88 & 66.85 $\pm$ 11.03 \\
\midrule
\textbf{Full Model} 
& \textcolor{forestgreen}{3.32 $\pm$ 2.51} & \textcolor{forestgreen}{0.40 $\pm$ 0.28} & \textcolor{forestgreen}{17.31 $\pm$ 3.37} & \textcolor{forestgreen}{85.69 $\pm$ 10.21} 
& \textcolor{forestgreen}{0.92 $\pm$ 0.59} & \textcolor{forestgreen}{0.70 $\pm$ 0.31} & \textcolor{forestgreen}{17.56 $\pm$ 2.18} & \textcolor{forestgreen}{71.80 $\pm$ 10.84} \\
\bottomrule
\end{tabular}
}
\endgroup
\end{table*}

\section{Discussion}
\label{sec:discussion}

\begin{figure}
    \centering
    \includegraphics[width=1\linewidth]{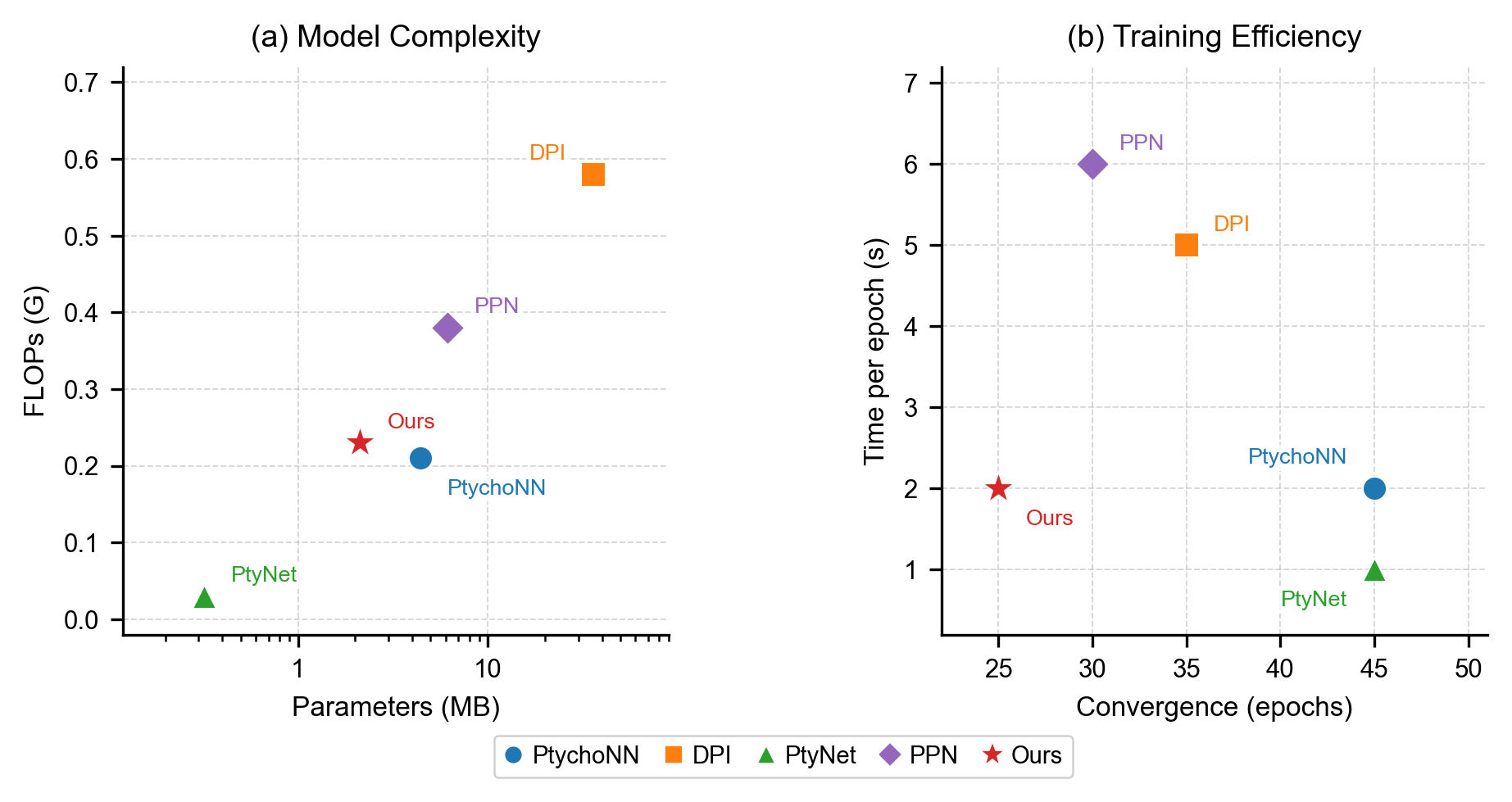}
    \caption{Model complexity and training efficiency comparison. (a) Computational complexity measured by FLOPs (G) versus model size (parameters in MB, log-scale on the horizontal axis). (b) Training efficiency measured by time per epoch (s) versus convergence epoch.}
    \label{fig:effi}
\end{figure}

The results consistently support the proposed circular phase representation and physics-informed constraints.
Several observations are worth noting.

\subsection{Circular versus direct phase prediction}
Representing phase on the unit circle avoids the $\pm\pi$ discontinuity inherent in direct regression on $(-\pi,\pi]$.
This is strongly supported by the ablation study: removing the circular representation (w/o Circ.\ Repr.) causes the largest degradation across both amplitude and phase metrics, together with much larger variance (e.g., amplitude SSIM std increases from $10.21\%$ to $24.26\%$).
This behavior is consistent with direct phase prediction, where small errors near the wrap boundary can produce large Euclidean gradients and unstable optimization.

\subsection{Synergy between representation and physical constraints}
The circular representation enables geometrically meaningful supervision.
The geodesic circular loss and manifold consistency constraint are defined on this representation, and the ablations show that both are important for phase structure.
Removing either term causes a large drop in phase SSIM, indicating that phase fidelity depends not only on representation, but also on enforcing the correct geometry during optimization.
The consistency loss remains useful even with explicit output normalization, as it regularizes decoder outputs before projection onto $S^1$.

\subsection{Dynamic-range handling in diffraction inputs}
Ptychographic diffraction patterns have extreme dynamic range, with weak high-angle fringes easily dominated by strong low-angle intensity.
The SADGS dual-gain design addresses this by providing complementary views of the same measurement.
The ablation results show that removing dual-gain scaling significantly increases amplitude error and degrades phase quality, confirming that exposing both strong and weak diffraction features improves reconstruction, especially for fine structures carried by weak fringes.
Extending to three gain levels brings only marginal benefit, suggesting that two gain views are already sufficient.

\subsection{Frequency-domain preservation and artifact suppression}
Spatial frequency analysis shows that the proposed method better preserves informative mid-to-high frequency components, especially in amplitude and mid-band phase variations.
This is consistent with the use of gradient- and SSIM-based supervision, which discourages over-smoothing and improves structural fidelity.
Circular phase modeling also reduces discontinuity-related optimization artifacts that may otherwise appear as spurious textures or excessive smoothing.

\subsection{Complexity and training efficiency}
Fig.~\ref{fig:effi} shows that these improvements do not require excessive computational cost.
In Fig.~\hyperref[fig:effi]{11(a)}, our model operates at a moderate FLOP budget while using substantially fewer parameters than heavier baselines such as DPI and PPN, giving a favorable quality--complexity trade-off.
Fig.~\hyperref[fig:effi]{11(b)} shows fast convergence (only \textbf{25} epochs) with low per-epoch time, resulting in substantially shorter total training time than DPI and PPN, and earlier convergence than PtychoNN and PtyNet.
This is consistent with the ablation results, which show that the main gains come from circular representation and geometric constraints rather than increased network capacity.

\section{Conclusion}
\label{sec:conclusion}

This paper presents a deep learning framework for ptychographic reconstruction that explicitly accounts for the circular geometry of phase, addressing the mismatch between periodic phase variables and standard Euclidean outputs and losses.
The method has two main components: (i) a circular phase representation based on cosine and sine outputs with explicit unit-circle normalization, which removes the $\pm\pi$ discontinuity and stabilizes phase learning; and (ii) a physics-informed composite loss that combines geodesic distance on $S^1$, a manifold consistency constraint, and gradient and SSIM terms for structure preservation.
These are integrated with a saturation-aware dual-gain design and lightweight decoders, yielding a favorable quality--efficiency trade-off.

Experiments show that the proposed method consistently outperforms existing state-of-the-art approaches, while additional results on experimental synchrotron data confirm robustness to real measurement effects.
Ablation studies further show that circular phase modeling and its associated constraints are the key contributors to the performance gains.
Future work will extend this framework to electron ptychography, where stronger phase wrapping, probe variation, partial coherence, and other electron-specific experimental effects pose additional challenges for stable and globally consistent reconstruction.

\section*{Acknowledgments}
This work was supported by the Australian Government through a Research Training Program (RTP) Scholarship. The authors also thank Han Yue for valuable discussions regarding PPN.

\bibliographystyle{IEEEtran}
\bibliography{references}

\end{document}